# Derivation of Tissue Properties from Basis-Vector Model Weights for Dual-Energy CT-Based Monte Carlo Proton Beam Dose Calculations

Maria Jose Medrano[1], Xinyuan Chen[2], Lucas Norberto Burigo[3], Joseph A. O'Sullivan[1], Jeffrey F. Williamson[4]

[1]Department of Electrical and Systems Engineering, Washington University in St. Louis, St. Louis, MO 63130

[2]Department of Biomedical Engineering, Washington University in St. Louis, St. Louis, MO 63130

[3]German Cancer Research Center (DKFZ), Heidelberg, Germany

[4]Department of Radiation Oncology, Washington University in St. Louis, St. Louis, MO 63110

**Objective:** We propose a novel method, basis vector model material indexing (BVM-MI), for predicting atomic composition and mass density from two independent basis vector model weights derived from dual-energy CT (DECT) for Monte Carlo (MC) dose planning.

**Approach:** BVM-MI employs multiple linear regression on BVM weights and their quotient to predict elemental composition and mass density for 70 representative tissues. Predicted values were imported into the TOPAS MC code to simulate proton dose deposition to a uniform cylinder phantom composed of each tissue type. The performance of BVM-MI was compared to the conventional Hounsfield Unit material indexing method (HU-MI), which estimates elemental composition and density based on CT numbers (HU). Evaluation metrics included absolute errors in predicted elemental compositions and relative percent errors in calculated mass density and mean excitation energy. Dose distributions were assessed by quantifying absolute error in the depth of 80% maximum scored dose (R80) and relative percent errors in stopping power (SP) between MC simulations using HU-MI, BVM-MI, and benchmark compositions. Lateral dose profiles were analyzed at R80 and Bragg Peak (RBP) depths for three tissues showing the largest discrepancies in R80 depth.

**Main Results:** BVM-MI outperformed HU-MI in elemental composition predictions, with mean RMSEs of 1.30% (soft tissue) and 0.1% (bony tissue), compared to 4.20% and 1.9% for HU-MI. R80 depth RMSEs were 0.2 mm (soft) and 0.1 mm (bony) for BVM-MI, vs. 1.8 mm and 0.7 mm for HU-MI. Lateral dose profile analysis showed overall smaller dose errors for BVM-MI across core, halo, and proximal aura regions

**Significance:** Fully utilizing the two-parameter BVM space for material indexing significantly improved TOPAS MC dose calculations by factors of 7 to 9 in RMS compared to the conventional HU-MI method demonstrating the potential of BVM-MI to enhance proton therapy planning, particularly for tissues with substantial elemental variability.

*Keywords:* dual-energy computed tomography, proton therapy, proton stopping power, Monte Carlo.

## 1. INTRODUCTION

Spectral X-ray CT has the potential to deliver highly accurate, patient-specific photon material information, crucial for improving dose calculations and reducing range uncertainties in proton radiotherapy (Bär *et al.*, 2017; Han *et al.*, 2017; Li *et al.*, 2017; Taasti *et al.*, 2018; Zhang *et al.*, 2018). Previous studies have shown that dual-energy CT (DECT) and multi-energy CT (MECT) material-decomposition strategies can reduce estimation errors for electron density ($\rho_e$), mean excitation energy (*I*-value), and SPR by 0.6%, 0.4%-1.3%, and 0.4-1.4%, respectively (Hünemohr *et al.*, 2014; Taasti *et al.*, 2016, 2018; Bär *et al.*, 2017; Zhang *et al.*, 2018). Given that most commercial planning tools accept only a single CT image, which is converted to stopping power ratio (SPR) using a 1D lookup table, previous DECT performance analyses have mostly focused on accurately predicting SPR images. This method significantly underutilizes the rich information provided by DECT and MECT, making it inadequate for analytical pencil-beam dose calculations which require atomic-number-sensitive mass-scattering powers in addition to SPR. Moreover, it is inadequate for accurate Monte Carlo dose-computation which requires detailed atomic composition and density information to model transport of secondary particles and to simulate elastic, inelastic, and nuclear proton scattering processes in complex geometries. However, obtaining a voxel-wise material decomposition breakdown from DECT and MECT is a challenging and ill-posed problem.

Various proposed segmentation, decomposition, and parametrization approaches have applied constraints to enable voxel-wise prediction of the atomic compositions and densities from DECT and MECT data (typically $\rho_e$ and $Z_{\text{eff}}$). In the segmentation method proposed by Landry et al., each voxel is classified by identifying the material composition that has the closest $\rho_e$ and $Z_{\text{eff}}$ values from a predefined library of materials (Landry *et al.*, 2013). Similarly, Malusek et al. proposed representing tissues as a linear combination of fundamental materials of known compositions (e.g., lipid, protein, and water), for which mass fractions are inferred from SECT or DECT measurements (Malusek *et al.*, 2013). More recent techniques have used principal component analysis (PCA) to generalize tissue decomposition by replacing predefined material components with orthogonal basis vectors derived from multi-energy CT data (Lalonde and Bouchard, 2016). Finally, parametrization-based methods model elemental mass fractions as continuous variables, which can be fitted to a database of human tissues with known compositions using one or more radiological quantities (Hünemohr *et al.*, 2014).

To utilize the capabilities of DECT in proton-therapy planning, our group previously proposed the Basis Vector Model (BVM) DECT material-decomposition method, which models photon cross sections as a weighted linear combination of bases materials (Williamson *et al.*, 2006; Han *et al.*, 2017). In earlier studies, we demonstrated that our BVM method supports highly accurate $\rho_e$, $Z_{\text{eff}}$, and *I*-value estimation, yielding SPR maps with subpercentage-uncertainty levels when integrated into a joint-statistical iterative model-based image reconstruction (JSIR-BVM) method (Zhang *et al.*, 2018; Medrano *et al.*, 2022). In this paper, we introduce the BVM Material-Indexing (BVM-MI) method, which allows for the continuous prediction of atomic composition and mass density using the two independent BVM weights derived from DECT imaging. Unlike previous segmentation, decomposition, and parametrization methods that utilize $\rho_e$ and $Z_{\text{eff}}$ to parametrize tissue elemental composition and density (Hünemohr *et al.*, 2014), our BVM-MI method directly employs our model's base-material components to maximize the 2D information derived from JSIR-BVM DECT image-reconstruction method. We compare the performance of our BVM-MI method against the conventional SECT material-indexing approach (HU-MI) (Schneider, 2000) in terms of predicted photon cross-sectional material properties and the shape and depth of the proton integrated depth dose (IDD) in controlled Monte Carlo simulations. The goal of our study is to evaluate the potential of BVM-MI to preserve BVM's subpercentage proton range uncertainty, emphasizing on the predicted elemental composition and resulting Monte Carlo dose distribution as the primary comparative endpoints, rather than a single dosimetric quantity like SPR.

## 2. METHODS AND MATERIALS

The following sections present *BVM material indexing (BVM-MI)*, a method for estimating the atomic composition and density of an unknown tissue from the accurate BVM-derived weights ($c_1$ and $c_2$). The accuracies of the resulting radiological quantities ($\rho_e$, *I*-value and SPR) and Monte Carlo dose distributions for 70 tissues (46 soft and 24 bony) of known elemental composition (Woodard and White, 1986; White, Woodard and Hammond, 1987) (see Supplementary Material S2) were compared to those of a standard SECT material-indexing approach (Schneider, 2000). Our BVM material indexing model was implemented and evaluated on Python 3.8.3 and Monte Carlo simulations were run on TOPAS 3.5 Toolkit.

### 2.1. Basis Vector Model

The photon linear attenuation coefficients of an arbitrary material in the diagnostic energy range can be represented by the BVM model as the linear combinations of two basis materials

$$\mu(\mathbf{x}, E) = c_1(\mathbf{x})\mu_1(E) + c_2(\mathbf{x})\mu_2(E), \tag{1}$$

Where $\mu_i$ are the linear attenuation coefficients of basis materials $i = 1:2$ for photon energy E and $c_i(\mathbf{x})$ denotes the basis material weight at image voxel location $x$. For this theoretical study, the basis materials polystyrene and a 23% aqueous solution of $CaCl_2$ were selected as they have shown to model linear attenuation coefficient of materials with atomic numbers of 2 to 20 with a 1-2% accuracy in the 20 keV to 1 MeV energy range (Williamson *et al.*, 2006). Accurate electron density ($\rho_e(\mathbf{x})$) and mean excitation energy ($I(\mathbf{x})$) maps can then be inferred from the basis material weights by,

$$\begin{aligned} \rho_e(\mathbf{x}) &= c_1(\mathbf{x})\rho_{e,1} + c_2(\mathbf{x})\rho_{e,2}, \\ I(\mathbf{x}) &= \exp(a \cdot r_c(\mathbf{x}) + b), \end{aligned} \tag{2}$$

where $\rho_{e,i}$ denotes the electron density of material $i$ and where a and b in (2) are derived by fitting *I*-values for tissues with known atomic composition to the electron density-weighted basis-component fraction

$$\hat{r}_c(\mathbf{x}) = \frac{\hat{c}_1(\mathbf{x})\rho_{e,1}}{\hat{c}_1(\mathbf{x})\rho_{e,1} + \hat{c}_2(\mathbf{x})\rho_{e,2}}, \tag{3}$$

which has previously been used as a surrogate for tissue composition in *I*-value calculations and has proved to be effective in estimating *I*-values with RMS errors of 2.36% (Zhang *et al.*, 2019).

The values of $c_1$, $c_2$, and $r_c$ for an arbitrary material can be derived by applying image and sinogram decomposition approaches on DECT data or directly through iterative reconstruction methods (O'Sullivan and Benac, 2007; Zhang et al., 2018). To test the BVM-MI in isolation from the entire reconstruction process, theoretical values of $c_1$, $c_2$, and $r_c$ for 70 reference tissues, with benchmark composition normalized to six major elements (hydrogen, carbon, nitrogen, oxygen, calcium, and phosphorus) were derived using the least-square approach outlined in previous studies (Zhang et al., 2019).

*2.2. BVM Material Indexing*

BVM-MI estimates an elemental composition of an arbitrary tissue, $y$, by a multiple linear regression model on the reconstructed JSIR-BVM basis-vector model weights, $c_{\{1,2\},y}$ and their corresponding weighted component ratio ($r_{c,y}$). To allow for the incorporation of low-density lung into the final parametrization, BVM material indexing uses a scaled version of the basis-material weights ($c'_{\{1,2\},y}$) in the final elemental parametrization of tissue $y$. The parameters $c'_{\{1,2\},y}$ can be defined as $c'_{\{1,2\},y} = c_{\{1,2\},y} \cdot \frac{\rho_{e,w}}{\rho_{e,y}}$ where $\rho_{e,w}$ and $\rho_{e,y}$ are the predicted electron densities of water and tissue $y$, and $\rho_{e,1}$ and $\rho_{e,2}$ are the electron densities of basis material $c_{1,y}$ and $c_{2,y}$, respectively. The scaled weighted component ratio for $c'_{\{1,2\},y}$ is equivalent to our original unscaled $r_{c,y}$; therefore, the final linear relationship between the scaled JSIR-BVM reconstructed basis vector model weights and weighted component ratio and a material's mass fraction for each element $k$ can be expressed as

$$w_{k,x} = \alpha_{1,k} c'_{1,y} + \alpha_{2,k} c'_{2,y} + \alpha_{3,k} r_{c,y} + \alpha_{4,k}, \tag{4}$$

where $k = 1:6$ corresponds to the six most abundant elements found in biological tissues (hydrogen, carbon, nitrogen, oxygen, phosphorus, and calcium), and $c'_{1,y}$, $c'_{2,y}$, and $r_{c,y}$ denote the scaled estimated BVM weights and component-weight ratio, respectively. Final $\alpha$ parameters were calculated by performing a multiple linear regression fit between theoretical mass fractions ($w_p$) of elements $p$ and benchmark $c'_1$, $c'_2$, and $r_{c,y}$ for 70 reference tissues of known material composition (Supplementary Material S2). Thyroid was not included in our comparison as it has previously shown large BVM modelling errors due to its high iodine content ($Z = 53$) (Han et al., 2017). Due to compositional differences between bony and soft tissues, particularly in phosphorus and calcium content, two sets of linear regression models (shown in Figure 1) were trained: one for bony tissues ($r_{c,y} < 0.75$) and one for soft tissues ($r_{c,y} \geq 0.75$). The predicted mass fractions of the six elements for each tissue were normalized to sum to 1. Using the estimated electron densities from Equation (2) and the predicted elemental composition, mass densities were calculated by applying the atomic mixture rule to the predicted weight fractions and elemental densities from the BVM-MI model.

*2.3. Performance Evaluation and Monte Carlo Range Studies*

BVM-MI was compared against the HU material indexing method from Schneider et al., which has been commonly used in MC simulation interfaces. The CT numbers (HU) at 120 kVp for the 70 tissues evaluated in our study were obtained from Schneider et al.'s original paper. Further details on our implementation of the HU-MI method can be found on Supplementary Material S6.

The accuracy of HU-MI and BVM-MI was evaluated by comparing: (1) discrepancies between radiological quantities (electron density, $I$-value, and stopping power) derived from predicted elemental compositions and their corresponding reference values, and (2) differences in proton-beam Monte Carlo dose distributions based on HU-MI, BVM-MI, and benchmark elemental compositions and densities. To calculate the $\rho_e$ and $I$-value corresponding to HU-MI and BVM-MI, we applied the mass density ($\rho_m$) and elemental mass fraction, $w_p$ from each indexing method to the atomic mixture and Bragg additivity rules per ICRU Report 37 (ICRU, 1984). The stopping power of each material for 200 MeV protons was calculated by the Bethe-Bloch equation without density and shell corrections, which are negligible for the clinically relevant energy range (ICRU, 1993). Estimated values of $\rho_m$, $I$-value, and SP were finally compared to their corresponding references computed from ground-truth densities and elemental composition. The percentage error in predicted elemental mass-fraction of the two MI methods was also assessed.

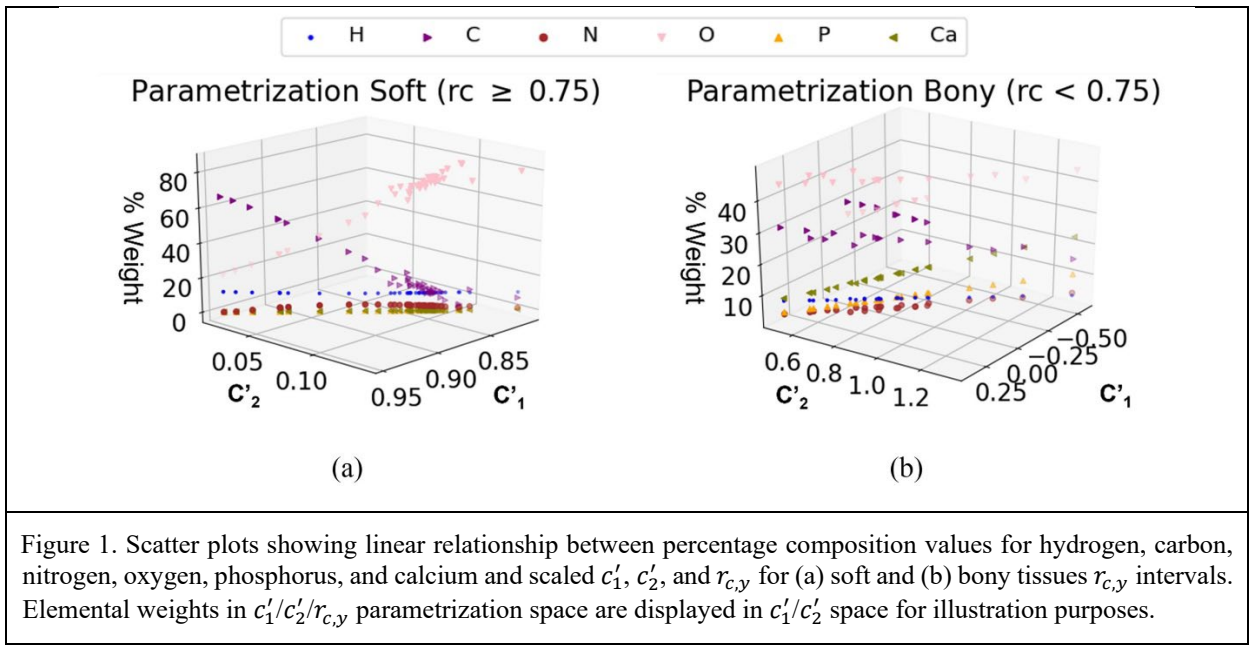

Figure 1. Scatter plots showing linear relationship between percentage composition values for hydrogen, nitrogen, oxygen, phosphorus, and calcium and scaled $c'_1$, $c'_2$, and $r_{c,y}$ for (a) soft and (b) bony tissues $r_{c,y}$ intervals. Elemental weights in $c'_1/c'_2/r_{c,y}$ parametrization space are displayed in $c'_1/c'_2$ space for illustration purposes.

To evaluate the generalizability of our BVM-MI model, we performed 5-fold cross-validation for each of the six elemental parametrizations. For each parametrization, the bony and soft tissue sets were divided into five folds. In each iteration, one fold was used for testing, and the remaining four folds were used for training. This process continued until all five folds were tested. The average percentage point error across all iterations was reported.

Our Monte Carlo analysis was performed by importing the HU-MI, BVM-MI, and ground truth densities, along with the mass fractional atomic compositions of the 70 evaluated tissues, into TOPAS Version 3.5, which employs the same physics models and interaction cross-sections as Geant4 10.06.p01. All simulations used the default Geant4 physics lists: g4em-standard_opt4, g4h-phy_QGSP_-BIC_HP, g4decay, g4ion-binarycascade, g4h-elastic_HP, and g4stopping. For each indexing method and material combination, TOPAS was used to compute IDD of a 200 MeV Gaussian pencil beam ($\sigma = 4$ mm) consisting of $2\times10^5$ particles incident on a uniform cylindrical phantom of 50 cm radius and 30 cm length. From each IDD, its range at 80% of its maximum dose (R80) served as the final endpoint. For inflated lung, the incident beam energy was reduced to 75

MeV to ensure that the IDD curve fit within the cylinder length. The dose to the medium was scored along the beam axis using 800 depth bins, each 0.038 mm thick in the cylinder's longitudinal direction. Each annular bin had a radius of 50 cm, with a single azimuthal bin covering 360 degrees. The standard deviations about the scored dose ranged from 0.2% to 0.5% in the [0, R80] depth range. The IDD curve was fitted to the Bortfeld function (Bortfeld, 1997), and the corresponding R80 of the proton beam was calculated from the fit.

To further investigate the impact of different material-indexing methods on our MC simulations, we hypothesized that tissue-composition errors would influence the proton beam's halo and aura doses due to variations in energy losses from neutron elastic scattering, particularly with respect to hydrogen, oxygen, and carbon concentrations in soft tissue, and calcium content in bony tissue. Pedroni et al. demonstrated that improper modeling of aura and halo doses can cause up to 10% dose errors when all pencil beams are summed (Pedroni *et al.*, 2005). To test this, additional simulations were conducted to evaluate discrepancies in the lateral spatial spread of the pencil beam, caused by multiple Coulomb scattering, inelastic scattering, and secondary neutral particle transport (gamma rays and neutrons). The lateral dose profiles of the proton beam at the Bragg peak and R80 depth were scored on a solid cylinder composed of one of the three representative soft or bony tissues that exhibited the largest R80 discrepancy between Monte Carlo simulations with benchmark material compositions and with HU-MI and BVM-MI-derived elemental compositions. Dose was scored in 500 annular bins, each 0.1 cm thick radially and axially, with each azimuthal bin covering 360 degrees, within a cylinder of 50 cm radius and 30 cm length. The dose to the annular bins at the R80 depth (R80) and the depth of the Bragg peak (RBP) were used to analyze lateral spatial spread characteristics.

To compare the lateral spread profiles from the BVM-MI and HU-MI MC simulations, the R80 and RBP bin depths were chosen so as to align with their corresponding positions in the ground truth MC simulations. A 3-bin boxcar averaging method was then used to smooth each profile to mitigate Monte Carlo statistical errors. The accuracy of the proton beam spread was quantified by calculating the overall mean absolute percent dose error in the lateral profiles along the beam's core, halo, and proximal aura regions between benchmark and HU-MI and BVM-MI MC simulations. To assess if differences between BVM-MI and HU-MI mean absolute percent dose errors were statistically significant, we tested the normality of the percent dose error distributions using the Shapiro-Wilk test. If the p-value from the normality test was greater than 0.05 (indicating normal distribution), we applied the paired t-test to evaluate the difference in the mean absolute percent error between the models. In cases where the p-value from the normality test was less than 0.05 (indicating non-normal distribution), we used the Wilcoxon signed-rank test, a non-parametric method, to evaluate the differences. For all statistical tests, a p-value less than 0.05 was considered statistically significant.

Following Gottschalk and Verbeek et al., we defined the core, halo, and aura regions as the lateral profile segments with doses between 100% to 10%, 10% to 0.01%, and below 0.01% of the maximum dose, respectively (Gottschalk *et al.*, 2015; Verbeek *et al.*, 2021). For our analysis, we evaluated errors within the core region (100% to 10%), the halo region (subdivided into proximal (10% to 1%) and distal (1% to 0.1%) components), and the proximal aura region (0.1% to 0.01%), which we define as the transitional area between the halo and aura regions. Analysis of percent error of doses below 0.01% was not feasible due to their high noise, so we focused on the proximal aura region, where the doses remained large enough to allow for statistically significant comparisons.

## 3. RESULTS

### 3.1. Predicted Elemental Fractions and Material Properties

Table I summarizes the resulting α values for the BVM-MI soft and bony tissue parametrization, as well as the overall accuracy of the HU and BVM-MI methods in predicting the elemental weight fractions for soft and bony tissues expressed in root-mean-square percentage point error. Overall, BVM-MI demonstrated superior accuracy compared to the conventional HU-MI method. The most significant differences were observed in the estimation of carbon and oxygen content, where the RMSEs for HU-MI and BVM-MI were 7.4/8.0% and 1.7/2.4% for soft tissue, and 4.8/4.9% and 0.3/0.3% for bony tissue, respectively. The RMSE from 5-fold cross-validation provided an estimate of the out-of-sample error, with a mean RMSE of 1.8% for soft tissue and 0.2% for bony tissue, both superior to the HU-MI method's mean RMSEs of 4.2% for soft tissue and 1.9% for bony tissue. Detailed scatter plots comparing the estimated and benchmark elemental fractions for the 70 investigated tissues are shown in Figure S1 of the Supplementary Material. Overall, the BVM-MI scatter plot results shown in Figure S1 for the five elements of interest in soft and bony tissues follow the identity line more closely than the HU-MI results, indicating better correspondence between the predicted and benchmark values. The greatest spread was observed in the scatter plot for nitrogen, with the highest discrepancy in nitrogen elemental composition being 2.8 percentage points. Figure 3 displays the percent error differences between benchmark and predicted $\rho_m$, I-value, and SP for the two MI methods and for each of the 46 and 24 soft and bony tissues. BVM-MI showed better correspondence to benchmark material properties for soft/bony tissues, with RMS $\rho_m$, I-value, and SP percent errors of 0.29/0.05%, 0.28/0.09%, and 0.10/0.03%, compared to the HU-MI method, which had errors of 0.77/0.34%, 1.90/1.51%, and 0.76/0.34%, respectively. The x-axis indices for each soft and bony tissue in Figures 2(a-f) are provided in Supplementary Material S2.

TABLE I. Fits of BVM material indexing applied to benchmark fractional elemental composition of 70 reference soft and bony tissues, along with corresponding root-mean-square (RMS) and maximum percentage point errors between predicted elemental compositions and benchmark values for the two investigated methods.

| | Soft | | | | Bony | | | | | |
|---|---|---|---|---|---|---|---|---|---|---|
| | H | C | N | O | H | C | N | O | Ca | P |
| $\alpha_{1,k}$ | -1721 | -39609 | 8469 | 35825 | -402 | -44366 | 1060 | 47422 | -3860 | 145 |
| $\alpha_{2,k}$ | 25895 | -184090 | -92373 | 263050 | -1505 | -48821 | 5523 | 46779 | -1582 | -395 |
| $\alpha_{3,k}$ | 23467 | 2034 | -85934 | 186510 | -871 | 2034 | 3615 | -6700 | 2402 | -480 |
| $\alpha_{4,k}$ | -21784 | 41085 | 77714 | -221240 | 1272 | 41085 | -4642 | -39299 | 1345 | 339 |
| $R^2$ | 0.51 | 0.99 | 0.45 | 0.98 | 0.99 | 0.99 | 0.93 | 0.99 | 0.99 | 0.99 |
| RMSE/Max Percentage Point Error | | | | | | | | | | |
| BVM Material Indexing RMSE/Max | 0.3/0.9 | 1.7/5.1 | 0.95/2.8 | 2.4/7.3 | 0.04/0.1 | 0.3/0.8 | 0.1/0.3 | 0.3/0.9 | 0.04/0.1 | 0.1/0.1 |
| BVM Material 5-Fold Cross- | 0.4 | 2.6 | 1.1 | 3.0 | 0.04 | 0.4 | 0.2 | 0.4 | 0.1 | 0.1 |

| Validation RMSE | | | | | | | | | | |
|---|---|---|---|---|---|---|---|---|---|---|
| HU Material Indexing RMSE/ Max | 0.3/0.8 | 7.4/28.4 | 1.1/4.0 | 8.0/31.0 | 0.2/0.5 | 4.8/9.4 | 0.5/0.8 | 4.9/9.6 | 0.5/1.2 | 0.2/0.5 |

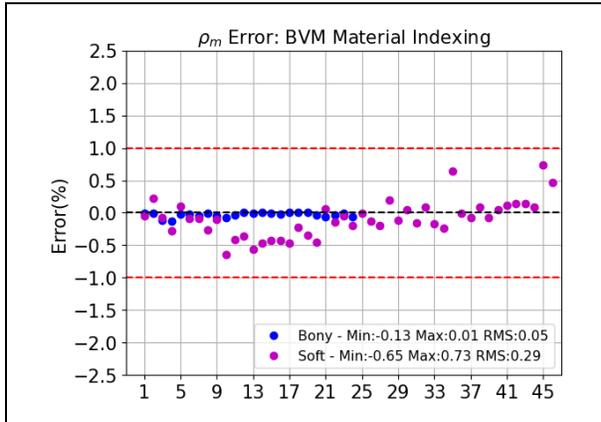 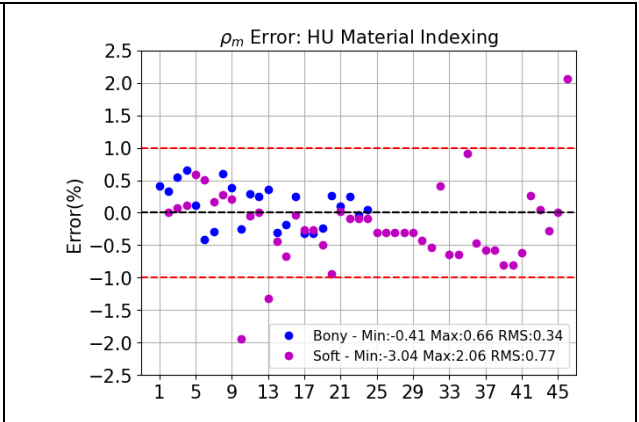

(a)          (b)

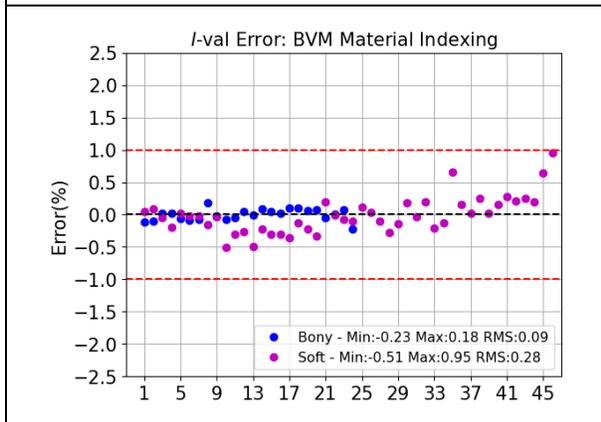 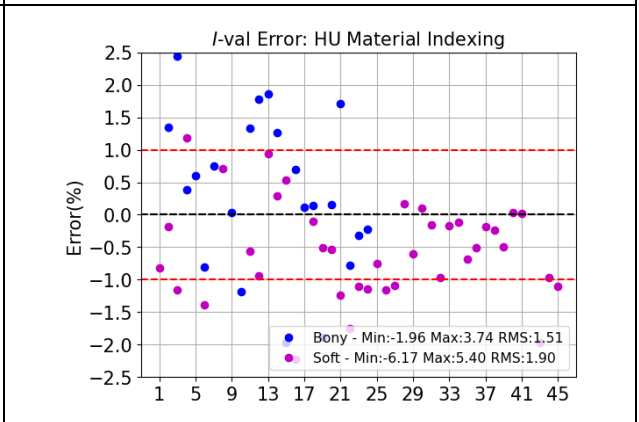

(c)          (d)

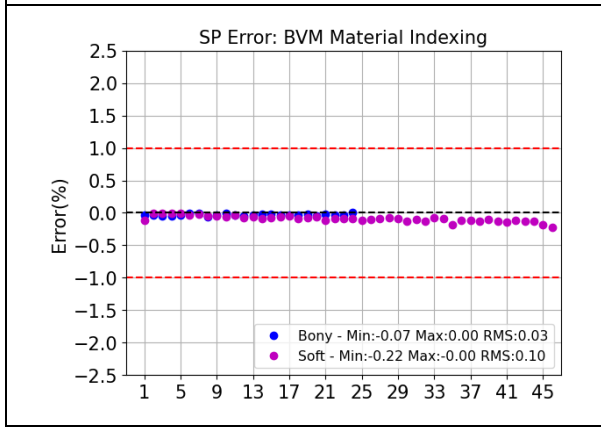 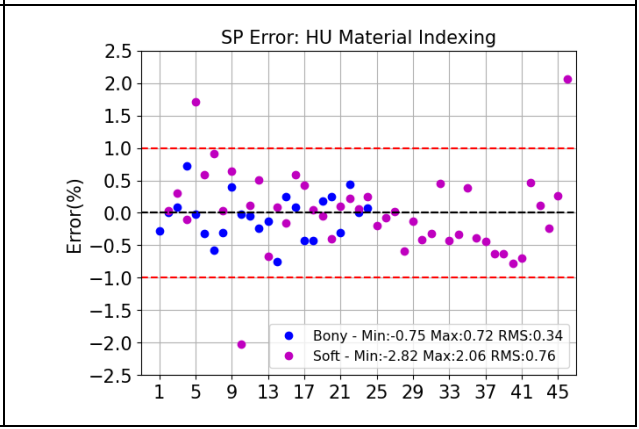

| (e) | (f) |

Figure 2. Overall discrepancies between predicted and benchmark $\rho_m$, $I$-value, and SP tissue parameters for BVM ((a), (c), and (e)) and HU ((b), (d), (f)) material-indexing methods. The correspondence between the x-axis integers and the 46 soft and 24 bony tissues is given by Table S2 in the Supplementary Material file. The black dotted line represents perfect correspondence, and the red dotted lines enclose the -1% to 1% error interval.

## 3.1. Monte Carlo Results

Figure 3 summarizes the discrepancy in estimated R80 range between benchmark and BVM-MI and HU-MI MC simulations for 46 soft and 24 bony tissues. BVM-MI provides more accurate R80 localization than HU-MI, with soft and bony RMS errors of 0.24 mm and 0.07 mm for BVM-MI, compared to 1.63 mm and 0.68 mm for HU-MI, respectively. The soft tissues with the largest R80 estimation errors for HU-MI were Brain Cerebrospinal Fluid, Cartilage, and Inflated Lung, while the largest R80 estimation errors for BVM-MI occurred in Cartilage, Eyelens, and Connective Tissue. For bony tissues, the largest R80 errors for HU-MI were observed in Femurtotalbone, Vertcolwhole, and Femurconicaltrochanter, while BVM-MI showed the largest errors in C4inclcartilagemale, Vertcolwhole, and D6L3inclcartilagem. The x-axis indices for each soft and bony tissue in Figures 3(a-b) are provided in Supplementary Material S2.

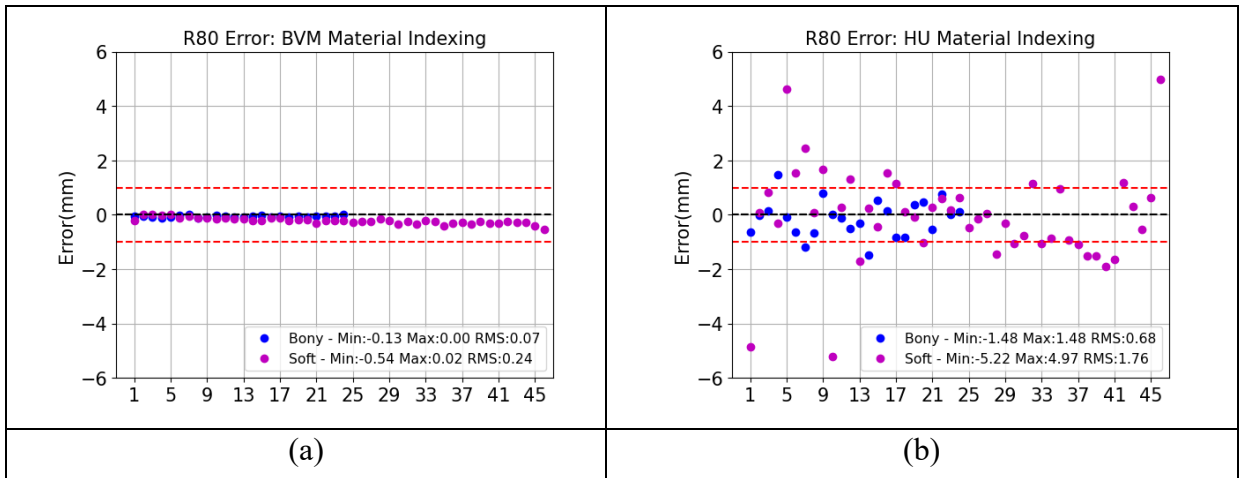

| (a) | (b) |

Figure. 3. Overall discrepancy between R80 corresponding to BVM and HU material indexing method and GT. The x-axis denotes each of the 46 and 24 soft and bony tissues in the reference material list. Black dashed line represents perfect correspondence, and the red dashed lines enclose -1 to 1 mm difference interval.

The two soft tissues with the highest R80 discrepancy (Brain Cerebrospinal Fluid and Cartilage) and the bony tissue with the highest R80 discrepancy (Vertebral Column Whole) were selected to investigate the effect of discrepancies in estimated hydrogen, carbon, oxygen, and calcium content on dose percentage errors in the core, halo, and aura regions. Figure 4 illustrates the IDD dose profiles for the three tissues selected for analysis, showing differences in R80 between ground truth (black line) and HU material indexing (blue line), which ranged from 1 to 5 mm. Notably, among the 70 tissues analyzed, Brain Cerebrospinal Fluid and Vertebral Column Whole also exhibited the highest percentage errors in carbon and oxygen content (Brain Cerebrospinal Fluid) and calcium content (Vertebral Column Whole).

As summarized in Table III, the greatest discrepancies in predicted elemental composition occurred for the soft tissues, Brain Cerebrospinal Fluid and Cartilage, where significant differences were observed in carbon and oxygen content. For Brain Cerebrospinal Fluid, carbon content was 0% (ground truth), 3% (BVM-MI), and 28% (HU-MI), while oxygen content was 89% (ground truth), 84% (BVM-MI), and 58% (HU-MI). Cartilage showed moderate differences in carbon content (10%, 7%, and 21% for ground truth, BVM-MI, and HU-MI, respectively) and oxygen content (76%, 80%, and 62% for ground truth, BVM-MI, and HU-MI, respectively). In comparison, Vertebral Column Whole exhibited relatively small discrepancies in calcium content, with values of 11%, 11%, and 12% for ground truth, BVM-MI, and HU-MI, respectively.

TABLE III. Hydrogen, carbon, and oxygen content, mass density ($\rho_m$) and mean excitation energy ($I$) for three representative soft and bony tissues with the highest R80 discrepancy. Calcium content (element with the highest atomic weight contribution for bony tissues) is also reported

|  | Brain Cerebrospinal Fluid | | | | | Cartilage | | | | | Vertebral Column Whole | | | | | |
| --- | --- | --- | --- | --- | --- | --- | --- | --- | --- | --- | --- | --- | --- | --- | --- | --- |
|  | H | C | O | $\rho_m$ $\left[\frac{g}{cm^3}\right]$ | $I$ [eV] | H | C | O | $\rho_m$ $\left[\frac{g}{cm^3}\right]$ | $I$ [eV] | H | C | O | Ca | $\rho_m$ $\left[\frac{g}{cm^3}\right]$ | $I$ [eV] |
|  | [%] | | | | | [%] | | | | | [%] | | | | | |
| GT | 11.2 | N/A | 88.8 | 1.01 | 75.3 | 9.8 | 10.1 | 75.7 | 1.10 | 76.9 | 7.2 | 26.0 | 47.5 | 10.6 | 1.3 | 86.5 |
| BVM-MI | 10.6 | 3.21 | 84.1 | 1.02 | 75.7 | 10.4 | 7.1 | 79.6 | 1.09 | 76.2 | 7.1 | 26.8 | 46.6 | 10.6 | 1.3 | 86.5 |
| HU-MI | 10.6 | 28.4 | 57.8 | 1.03 | 71.2 | 9.4 | 20.7 | 62.2 | 1.08 | 74.7 | 7.1 | 33.5 | 38.7 | 11.7 | 1.3 | 86.1 |

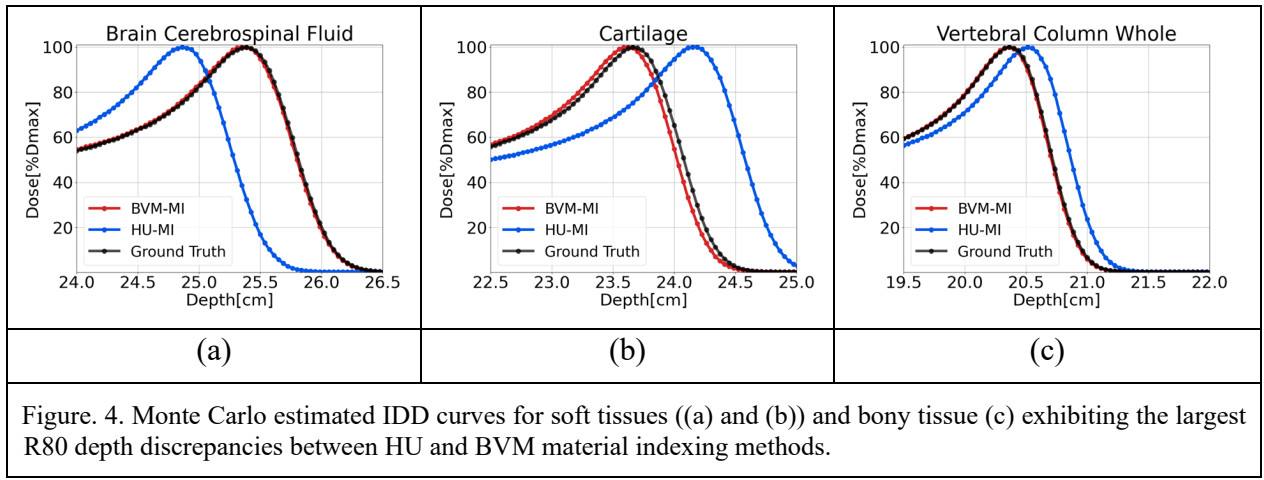

Figure. 4. Monte Carlo estimated IDD curves for soft tissues ((a) and (b)) and bony tissue (c) exhibiting the largest R80 depth discrepancies between HU and BVM material indexing methods.

The lateral dose profiles for the three representative tissues at the R80 and RBP proton beam depths are shown in Figure S2 of the Supplementary Materials. Table S1 summarizes the mean absolute percent error in dose differences between the benchmark and BVM-MI, and HU-MI MC simulations at the RBP proton beam depth, for the core, halo, and aura regions. Table IV provides a summary of the mean absolute percent error in dose differences for the core, halo, and proximal aura regions between benchmark and BVM-MI and HU-MI MC simulations at the R80 proton beam depth. For Brain Cerebrospinal Fluid, statistically significant differences in mean absolute percent error between BVM-MI and HU-MI were observed at both RBP and R80 depths, with BVM-MI showing smaller errors. At the R80 depth, the errors for BVM-MI ranged from 3%-16% compared to 11%-89% for HU-MI, and at the RBP depth, the errors ranged from 2%-53% for BVM-MI and 6%-56% for HU-MI. For cartilage, statistically significant differences in mean absolute percent error between BVM-MI and HU-MI were found at the R80 depth in the core and halo regions, and at the RBP depth in the halo region. Percent error differences at the R80 depth ranged from 9%-48% for BVM-MI and 24%-43% for HU-MI, while at the RBP depth, the differences ranged from 3%-55% for BVM-MI and 2%-41% for HU-MI. For Vertebral Column Whole, no statistically significant differences were observed at the RBP or R80 depth, indicating similar performance between BVM-MI and HU-MI. At the R80 depth, mean absolute percent error differences ranged from 11%-42% for BVM-MI and 11%-46% for HU-MI.

TABLE IV. Mean Absolute Percent Error Between Doses in Core, Halo, and Proximal Aura Regions for Benchmark, BVM-MI, and HU-MI MC simulations scored at R80 proton beam range. An asterisk denotes statistically significant differences ($p < 0.05$) between BVM-MI and HU-MI results.

|  |  | Brain Cerebrospinal Fluid | | Cartilage | | Vertebral Column Whole | |
|---|---|---|---|---|---|---|---|
|  |  | BVM-MI | HU-MI | BVM-MI | HU-MI | BVM-MI | HU-MI |
| Core | $D_\% \geq 10\%$ | 3.0%* | 11.3% | 8.5%* | 24.4% | 10.6% | 10.5% |
| Halo | $1\% \leq D_\% < 10\%$ | 3.7%* | 30.4% | 10.7%* | 29.8% | 6.8% | 13.1% |
|  | $0.1\% \leq D_\% < 1\%$ | 10.0%* | 79.5% | 11.7%* | 21.8% | 19.3% | 17.6% |
| Proximal Aura | $0.01\% \leq D_\% < 0.1\%$ | 15.9%* | 89.0% | 48.1% | 42.8% | 41.8% | 46.3% |

## 4. DISCUSSION

Unlike previously proposed DECT material indexing methods that rely on estimated $\rho_e$-$Z_{\text{eff}}$, our scheme uses BVM basis-weight material decomposition to directly estimate elemental tissue content. The BVM material decomposition method not only enhances the accuracy of photon cross-section modeling in the diagnostic X-ray range but also offers the added benefit of compatibility with polychromatic iterative dual-energy CT reconstruction processes, such as JSIR-BVM (Han *et al.*, 2017). The continuous estimation of elemental tissue content through BVM-MI leverages the improved material decomposition from BVM to enable the first full integration of basis-material weight images from JSIR-BVM reconstructions into Monte Carlo simulations for proton-beam treatment planning. Previous simulation and experimental studies have demonstrated that JSIR-BVM predicts $\rho_e$, *I*-value, and SPR of human tissues with RMS errors of 0.13%, 1.55%, and 0.16% (Zhang *et al.*, 2018) and has the potential to reduce range uncertainties to subpercentage levels (Medrano *et al.*, 2022). In this study, the BVM-MI method demonstrated RMSE accuracies of 0.10% for soft tissues and 0.03% for bony tissues in predicting SP. These results are comparable to or better than the theoretical accuracy of our original BVM approach, which predicts SP directly from the basis weights, with errors of 0.35% for soft tissues and 0.17% for bony tissues (Zhang *et al.*, 2018). Consequently, the integration of BVM-MI into our JSIR-BVM reconstruction process is expected to preserve the subpercentage range uncertainties previously demonstrated for patient DECT scans acquired on commercial scanners (Medrano *et al.*, 2022).

We observe that BVM-MI significantly outperforms the traditional SECT method in estimating carbon and oxygen mass fractions for both soft and bony tissues, leading to substantial improvements in the estimated *I*-value (RMSE of 0.3%/0.1% for soft/bony tissues for BVM-MI vs. 2.2%/1.3% for HU-MI). This improvement is due to BVM-MI's ability to more effectively capture tissue-to-tissue variability in carbon and oxygen content, which is influenced by variations in water, protein, and lipid mass fractions. These variations, combined with the distinct elemental $I$-values of carbon (81 eV) and oxygen (106 eV), significantly impact $I$-value prediction, making precise estimation of oxygen and carbon mass fractions critical for accurate $I$-value results. Through this study, we also found that the direct calculation of mean excitation values using elemental compositions predicted by the SECT parameterization led to slightly more accurate $I$-values (1.9%/1.5% for soft and bony tissues) than those predicted by our previously proposed BVM and $\rho_e$-$Z_{eff}$ DECT models (2.2%/1.3% for soft and bony tissues). However, DECT models still provided more accurate $\rho_e$ predictions (errors of 0.1%/0.03% and 0.8%/0.3% for soft/bony tissues using DECT and SECT-based methods, respectively), resulting in more accurate SPR estimates of 0.4% and 0.2% for soft and bony tissues. Additionally, the BVM-MI method demonstrated the potential to fully utilize the BVM information contained in each ($c_1$, $c_2$, and $r_c$) voxel-specific vector to generate accurate $\rho_e$, SPR, and $I$-values, surpassing the performance of HU-MI and our previous BVM approach with direct $I$-value prediction. The resulting BVM-MI stopping power RMSE is comparable to that of a previously proposed DECT-based continuous $\rho_e/Z_{eff}$-MI parametrization method, with 0.2% and 0.1% errors for soft and bony tissue, respectively (Hünemohr et al., 2014). BVM-MI's soft and bony mass density predictions (0.3% and 0.1%) outperformed the HU-MI method (0.8% and 0.3%) and were comparable to those reported for the Huhnemohr model (0.2% and 0.1%) (Hünemohr et al., 2014). Unlike the HU-MI $\rho_m$ predictions, which depend on five different linear fits to compensate for inherent CT number ambiguities and cover the entire tissue spectrum (Schneider, 2000), BVM-MI leverages the material characterization embedded in the BVM weights to directly derive $\rho_m$ from DECT spectral information via the JSIR-BVM algorithm. Furthermore, by proposing a normalization scheme for our basis vector model weights and weighted component ratio, BVM-MI bypasses the need for a special $\rho_e/\rho_m$ fitting interval to estimate mass density for low-density tissues, such as lung. Unlike previous studies, which required defining a special interval to determine a linear fit between HU and $\rho_m$ for SECT and $\rho_e$ and $\rho_m$ for DECT for low-density tissues, our approach allows the elemental composition of inflated lung, and consequently its mass density, to be directly predicted from a soft tissue parametrization fit.

The improved accuracy of the predicted BVM-MI mass density and elemental compositions directly contributed to the enhanced R80 proton beam path results in our MC simulations (Figure 3). When comparing the two material indexing methods across 70 representative tissues, our MC simulations showed R80 depth RMSE errors of 0.2 mm/0.1 mm for BVM-MI and 1.8 mm/0.7 mm for HU-MI, highlighting the superior performance of BVM-MI in predicting proton beam range. Additionally, the distribution of range differences observed in Figure 3 closely mirrors the distribution of stopping power errors shown in Figure 4. As noted by Hunemohr et al., this relationship arises from the fact that proton range is primarily determined by electronic energy loss at therapeutic proton energies, with nuclear interactions having a negligible effect (Hünemohr et al., 2014). Therefore, errors in proton range prediction are more significantly influenced by discrepancies in mass density, and thus electron density, than by variations in elemental composition, including carbon and oxygen mass fractions.

To better understand the overall impact of material composition in MC simulations, our study analyzed prediction errors in the core, halo, and aura regions, where nuclear interactions, sensitive to hydrogen, carbon, and oxygen content, can influence predictions in dose deposition, especially for tissues with substantial variation in elemental content between different material indexing methods. For our study, we selected three tissues with the highest R80 prediction errors for soft and bony tissues. As shown in Table III, discrepancies in the elemental content of the three tissues, specifically for the key elements involved in nuclear interactions, ranged from minimal (Vertebral Column Whole, with percentage errors ranging from 1-9%) to significant (Brain Cerebrospinal Fluid, with percentage errors ranging from 1-30%). These variations in elemental composition were reflected in the differences observed in the lateral spread of the proton beam at the core, halo, and aura regions. For Brain Cerebrospinal Fluid, discrepancies in predicted elemental composition led to statistically significant errors in mean absolute percent dose across different radial distances from the core to the aura region. These errors ranged from 3-16% and 2-53% for BVM-MI, and 2-53% and 6-56% for HU-MI at the RBP and R80 proton ranges, respectively. In contrast, no significant differences in dose error were observed between BVM-MI and HU-MI for Vertebral Column Whole at both RBP and R80 ranges, and only mild discrepancies were observed for Cartilage.

Although previous studies have analyzed the characterization of the core, halo, and aura regions by different Monte Carlo simulation platforms, to our knowledge, this is the first study to evaluate the sensitivity of proton beam range and dose-distribution to errors in material composition and density. Based on our results, we found that 0.2%-8% RMSE errors in HU-MI predicted material composition can lead to range errors of 1-5 mm. Furthermore, only the elemental composition differences for soft tissues were significant enough to cause major to mild errors in dose deposition across the core, halo, and aura regions. This discrepancy may be attributed to the different normalizations and compensations performed within the TOPAS interface. Future studies should further explore the limits of these compensations within TOPAS and investigate the sensitivity of simulated nuclear interactions to errors in material composition. The minor differences in lateral profiles predicted by BVM-MI and HU-MI compared to benchmark calculations might be surprising, especially given that Verbeek et al. showed that secondary neutrons contribute up to 25%-50% of the dose at off-axis distances of 3 to 8 cm (Verbeek *et al.*, 2021). However, this can be explained by the fact that elastic collisions between fast neutrons and hydrogen nuclei account for 70%-95% of their energy losses in soft tissue, and both BVM-MI and HU-MI predict hydrogen weight fractions with reasonable accuracy. Interestingly, BVM-MI performed well for Brain Cerebrospinal Fluid, despite the significant differences in its elemental composition, particularly in hydrogen, carbon, and oxygen content. For example, BVM-MI predicted carbon content (10.6%) and oxygen content (84.1%) with high accuracy, in contrast to HU-MI, which showed much larger discrepancies in carbon and oxygen relative to the ground truth. This suggests that BVM-MI may be more robust in predicting proton range and dose distribution in soft tissues, even those with significant elemental variability. Future studies will continue investigating the performance of BVM-MI and HU-MI under conditions of substantial soft tissue elemental variability.

An important next step is to test BVM-MI on basis-weight images reconstructed with our iterative JSIR-BVM method from raw sinogram data. Since TOPAS and other condensed-history transport code do not calculate material-specific libraries on the fly, a method for determining the optimal but discrete set of TOPAS materials is needed. One approach explored by other investigators, uses k-means clustering on voxel-wise reconstructed elemental compositions which involves developing an in-house TOPAS extension; however, this method still needs to be tested on multi-slice clinical data (Lalonde *et al.*, 2018). In addition, BVM-MI needs to be tested in more realistic simulation heterogeneous tissue phantoms. In this study, we evaluated the potential of using a

BVM material indexing method to derive the material properties (mass density and fractional elemental composition) of biological tissues from DECT data for accurate Monte Carlo based proton beam dose calculations. Besides its role in proton therapy treatment planning, BVM-MI could also be used in treatment verification by imaging secondary signals arising from proton-induced nuclear reactions (Enghardt *et al.*, 2004; Min *et al.*, 2006; Testa *et al.*, 2008; Landry *et al.*, 2013). This would be especially helpful in the Monte Carlo modelling of PET activity profiles where uncertainties in carbon and oxygen concentrations add up to 1 mm to range uncertainties (España and Paganetti, 2010). Other possible applications include dose calculations for other hadron therapy modalities, e.g., carbon ions, and classification of healthy and cancerous tissue by taking advantage of the higher hydrogen concentration in tumors (Brown M, 1969).

Despite its benefits, a limitation of the BVM material indexing method is that it relies on a two-parameter based calibration method attempting to estimate six material composition parameters. Lalonde et al. have previously shown the advantages of using multiple-energy computed tomography (MECT) for material characterization in proton-therapy dose calculations when CT images are corrupted by beam-hardening, partial volume, and streaking artifacts (Lalonde, Bär and Bouchard, 2017; Lalonde *et al.*, 2018). Hence, a valuable follow-up to our study would be to test the potential of JSIR-BVM in multi-energy CT reconstruction and its potential to more accurately estimate tissue elemental composition needed for Monte Carlo hadron therapy and brachytherapy dose calculations (Evans *et al.*, 2013). This could be achieved by adapting existing material PCA decomposition methods to JSIR-BVM basis weight images or even directly incorporating the PCA approach directly into our reconstruction process. Future studies will also focus on refining the BVM method to model thyroid tissue, which has previously shown high BVM modeling errors due to its high iodine content (Z=53), a limitation of the current BVM model (Han *et al.*, 2017). One potential solution is to apply the BVM model to multi-energy CT data, which provides more detailed attenuation information for iodine-containing tissues. Additionally, incorporating iodine as a basis material into the BVM model could improve its ability to predict elemental composition and density in iodine-rich tissues, such as the thyroid. Another promising direction is to explore deep learning techniques for directly learning the optimal $\alpha$ weights and deriving fractional elemental composition images from reconstructed $c_1'$, $c_2'$, and $r_c'$ maps (Lalonde and Bouchard, 2016; Lyu *et al.*, 2021). Shifting from a voxel-wise to image-wise analysis, deep learning could leverage geometric priors to enforce geometrical correlations among voxels, including organ types and boundaries (Fujiwara *et al.*, 2022). Given that reconstructed BVM weights are robust to noise and beam hardening artifacts, our BVM material indexing network has the potential to focus more effectively on learning the mapping to fractional elemental composition than other DECT mapping methods that are susceptible to noise in low- and high-energy CT input images (Zhang *et al.*, 2019).

## 5. CONCLUSIONS

Our proposed material-indexing scheme, BVM-MI, enables DECT statistical iterative reconstruction (SIR) algorithms to estimate elemental composition of tissue voxels, as well as radiological quantities such as SPR. This, in turn, enables our previously validated DECT JSIR-BVM reconstructions process to support accurate Monte Carlo as well as pencil-beam dose calculations. BVM-MI improves upon the accuracy of our prior BVM model for estimating densities, *I*-values, and stopping powers, for low (e.g., lung) and unit density tissues. Proton-beam Monte Carlo simulations demonstrate that BVM-MI contributes less than 0.2 mm to range uncertainties of 200 MeV protons in homogeneous phantoms. Thus BVM-MI preserves and potentially enhances the previously established subpercentage range-uncertainty achievable by JSIR-BVM DECT material-property imaging (Medrano *et al.*, 2022). Future studies will focus on further evaluating the BVM-MI method on JSIR-BVM reconstructed BVM weights and in more realistic clinical geometries and practical integration into TOPAS and other proton-therapy Monte Carlo codes.

## ACKNOWLEDGMENTS

This work was supported by R01 CA212638 and Imaging Sciences Pathway T32 EB014855 from the United States National Institutes of Health.

## REFERENCES

Bär, E., Lalonde, A., Royle, G., Lu, H.M. and Bouchard, H. (2017) 'The potential of dual-energy CT to reduce proton beam range uncertainties', *Medical Physics*, 44(6), pp. 2332–2344.

Bortfeld, T. (1997) 'An analytical approximation of the Bragg curve for therapeutic proton beams', *Medical Physics*, 24(12), pp. 2024–2033.

Brown M, P.PB. (1969) 'Neutron radiography in biologic media. Techniques, observations, and implications.', *The American journal of roentgenology, radium therapy, and nuclear medicine.*, 3(106), pp. 472–485.

Enghardt, W., Crespo, P., Fiedler, F., Hinz, R., Parodi, K., Pawelke, J. and Pönisch, F. (2004) 'Charged hadron tumour therapy monitoring by means of PET', *Nuclear Instruments and Methods in Physics Research, Section A: Accelerators, Spectrometers, Detectors and Associated Equipment*, 525(1–2), pp. 284–288.

España, S. and Paganetti, H. (2010) 'The impact of uncertainties in the CT conversion algorithm when predicting proton beam ranges in patients from dose and PET-activity distributions', *Physics in Medicine and Biology*, 55(24), pp. 7557–7571.

Evans, J.D., Whiting, B.R., O'Sullivan, J.A., Politte, D.G., Klahr, P.H., Yu, Y. and Williamson, J.F. (2013) 'Prospects for in vivo estimation of photon linear attenuation coefficients using postprocessing dual-energy CT imaging on a commercial scanner: Comparison of analytic and polyenergetic statistical reconstruction algorithms', *Medical Physics*, 40(12).

Fujiwara, D., Shimomura, T., Zhao, W., Li, K.W., Haga, A. and Geng, L.S. (2022) 'Virtual computed-tomography system for deep-learning-based material decomposition', *Physics in Medicine and Biology*, 67(15).

Gottschalk, B., Cascio, E.W., Daartz, J. and Wagner, M.S. (2015) 'On the nuclear halo of a proton pencil beam stopping in water', *Physics in Medicine and Biology*, 60(14), pp. 5627–5654.

Han, D., Porras-Chaverri, M.A., O'Sullivan, J.A., Politte, D.G. and Williamson, J.F. (2017) 'Technical note: On the accuracy of parametric two-parameter photon cross-section models in dual-energy CT applications', *Medical Physics*, 44(6), pp. 2438–2446.

Hünemohr, N., Paganetti, H., Greilich, S., Jäkel, O. and Seco, J. (2014) 'Tissue decomposition from dual energy CT data for MC based dose calculation in particle therapy', *Medical Physics*, 41(6).

ICRU (1984) *Stopping powers for electrons and positrons. Report No. 37*. Bethesda, MD: International Commission on Radiation Units and Measurements.

ICRU (1993) *Stopping powers and ranges for protons and alpha particles. Report No. 49*. Bethesda, MD: International Commission on Radiation Units and Measurements.

Lalonde, A., Bär, E. and Bouchard, H. (2017) 'A Bayesian approach to solve proton stopping powers from noisy multi-energy CT data', *Medical Physics*, 44(10), pp. 5293–5302.

Lalonde, A. and Bouchard, H. (2016) 'A general method to derive tissue parameters for Monte Carlo dose calculation with multi-energy CT', *Physics in Medicine and Biology*, 61(22), pp. 8044–8069.

Lalonde, A., Simard, M., Remy, C., Bär, E. and Bouchard, H. (2018) 'The impact of dual- and multi-energy CT on proton pencil beam range uncertainties: A Monte Carlo study', *Physics in Medicine and Biology*, 63(19).

Landry, G., Parodi, K., Wildberger, J.E. and Verhaegen, F. (2013) 'Deriving concentrations of oxygen and carbon in human tissues using single- and dual-energy CT for ion therapy applications', *Physics in Medicine and Biology*, 58(15), pp. 5029–5048.

Li, B., Lee, H.C., Duan, X., Shen, C., Zhou, L., Jia, X. and Yang, M. (2017) 'Comprehensive analysis of proton range uncertainties related to stopping-power-ratio estimation using dual-energy CT imaging', *Physics in Medicine and Biology*, 62(17), pp. 7056–7074.

Lyu, T., Zhao, W., Zhu, Y., Wu, Z., Zhang, Y., Chen, Y., Luo, L., Li, S. and Xing, L. (2021) 'Estimating dual-energy CT imaging from single-energy CT data with material decomposition convolutional neural network', *Medical Image Analysis*, 70, p. 102001.

Malusek, A., Karlsson, M., Magnusson, M. and Carlsson, G.A. (2013) 'The potential of dual-energy computed tomography for quantitative decomposition of soft tissues to water, protein and lipid in brachytherapy', *Physics in Medicine and Biology*, 58(4), pp. 771–785.

Medrano, M., Liu, R., Zhao, T., Webb, T., Politte, D.G., Whiting, B.R., Liao, R., Ge, T., Porras-Chaverri, M.A., O'Sullivan, J.A. and Williamson, J.F. (2022) 'Towards subpercentage uncertainty proton stopping-power mapping via dual-energy CT: Direct experimental validation and uncertainty analysis of a statistical iterative image reconstruction method', *Medical Physics*, 49(3), pp. 1599–1618.

Min, C.H., Kim, C.H., Youn, M.Y. and Kim, J.W. (2006) 'Prompt gamma measurements for locating the dose falloff region in the proton therapy', *Applied Physics Letters*, 89(18).

O'Sullivan, J.A. and Benac, J. (2007) 'Alternating minimization algorithms for transmission tomography', *IEEE Transactions on Medical Imaging*, 26(3), pp. 283–297.

Pedroni, E., Scheib, S., Böhringer, T., Coray, A., Grossmann, M., Lin, S. and Lomax, A. (2005) 'Experimental characterization and physical modelling of the dose distribution of scanned proton pencil beams', *Physics in Medicine and Biology*, 50(3), pp. 541–561.

Schneider, W.T.B. and W.S. (2000) 'Correlation between CT numbers and tissue parameters needed for Monte Carlo simulations of clinical dose distributions', *Phys. Med. Biol*, 45(2), pp. 459–478.

Taasti, V.T., Hansen, D.C., Michalak, G.J., Deisher, A.J., Kruse, J.J., Muren, L.P., Petersen, J.B.B. and McCollough, C.H. (2018) 'Theoretical and experimental analysis of photon counting detector CT for proton stopping power prediction', *Medical Physics*, 45(11), pp. 5186–5196.

Taasti, V.T., Petersen, J.B.B., Muren, L.P., Thygesen, J. and Hansen, D.C. (2016) 'A robust empirical parametrization of proton stopping power using dual energy CT', *Medical Physics*, 43(10), pp. 5547–5560.

Testa, E., Bajard, M., Chevallier, M., Dauvergne, D., Le Foulher, F., Freud, N., Ĺtang, J.M., Poizat, J.C., Ray, C. and Testa, M. (2008) 'Monitoring the Bragg peak location of 73 MeVu carbon ions by means of prompt γ-ray measurements', *Applied Physics Letters*, 93(9).

Verbeek, N., Wulff, J., Bäumer, C., Smyczek, S., Timmermann, B. and Brualla, L. (2021) 'Single pencil beam benchmark of a module for Monte Carlo simulation of proton transport in the PENELOPE code', *Medical Physics*, 48(1), pp. 456–476.

White, D.R., Woodard, H.Q. and Hammond, S.M. (1987) 'Average soft-tissue and bone models for use in radiation dosimetry', *The British Journal of Radiology*, 60(717), pp. 907–913.

Williamson, J.F., Li, S., Devic, S., Whiting, B.R. and Lerma, F.A. (2006) 'On two-parameter models of photon cross sections: Application to dual-energy CT imaging', *Medical Physics*, 33(11), pp. 4115–4129.

Woodard, H.Q. and White, D.R. (1986) 'The composition of body tissues', *The British Journal of Radiology*, 59(708), pp. 1209–1219.

Zhang, S., Han, D., Politte, D.G., Williamson, J.F. and O'Sullivan, J.A. (2018) 'Impact of joint statistical dual-energy CT reconstruction of proton stopping power images: Comparison to image- and sinogram-domain material decomposition approaches', *Medical Physics*, 45(5), pp. 2129–2142.

Zhang, S., Han, D., Williamson, J.F., Zhao, T., Politte, D.G., Whiting, B.R. and O'Sullivan, J.A. (2019) 'Experimental implementation of a joint statistical image reconstruction method for proton stopping power mapping from dual-energy CT data', *Medical Physics*, 46(1), pp. 273–285.

# SUPPLEMENTARY MATERIAL

# Derivation of Tissue Properties from Basis-Vector Model Weights for Dual-Energy CT-Based Monte Carlo Proton Beam Dose Calculations

## S1. Elemental Fractional Composition Error

Differences between benchmark and estimated elemental fractional composition for the 70 analyzed tissues are illustrated on the scatter plots in Figure S1. As can be observed, the BVM parametrization results very closely follow the black line which indicates perfect correspondence between predicted and benchmark values.

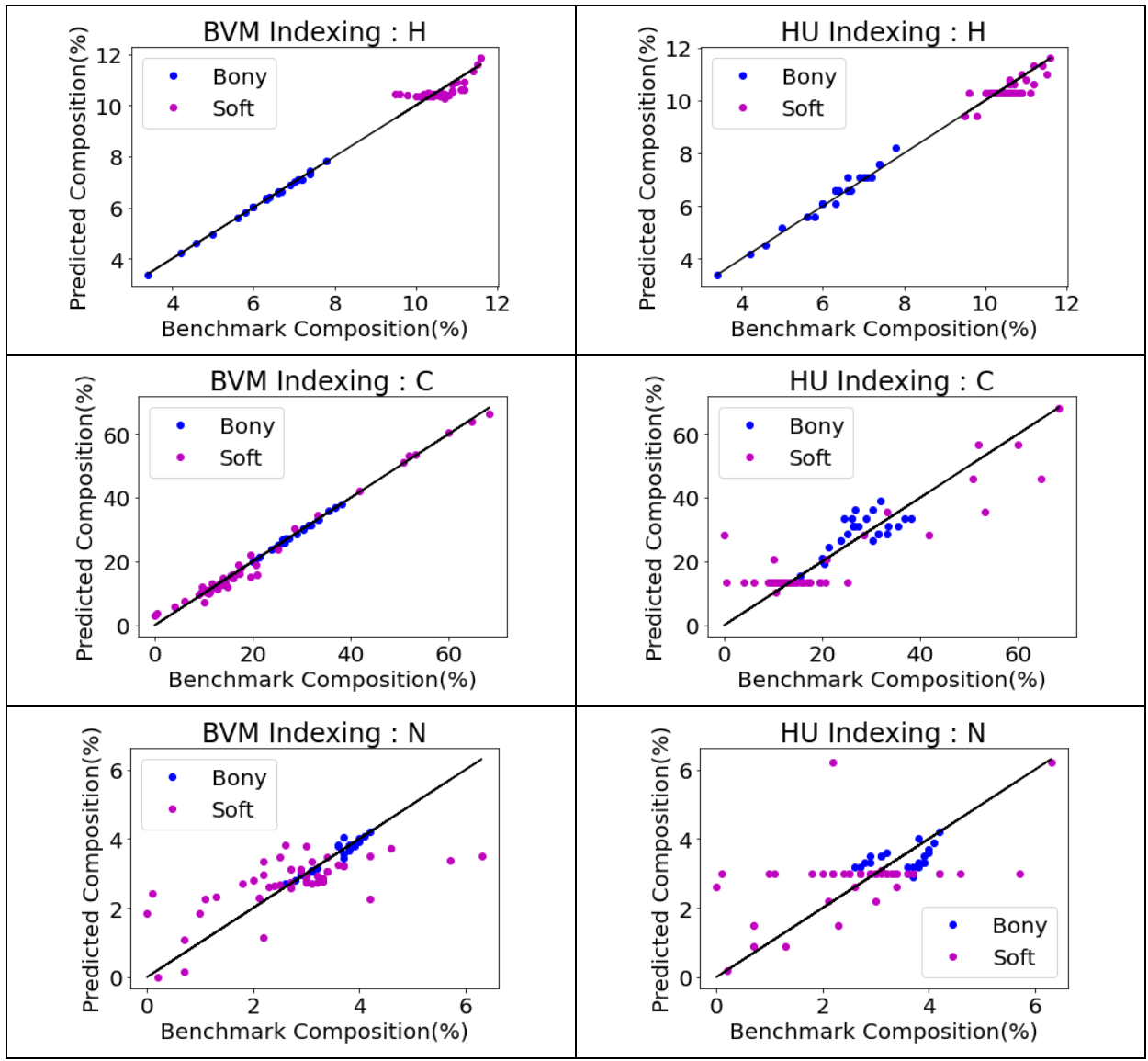

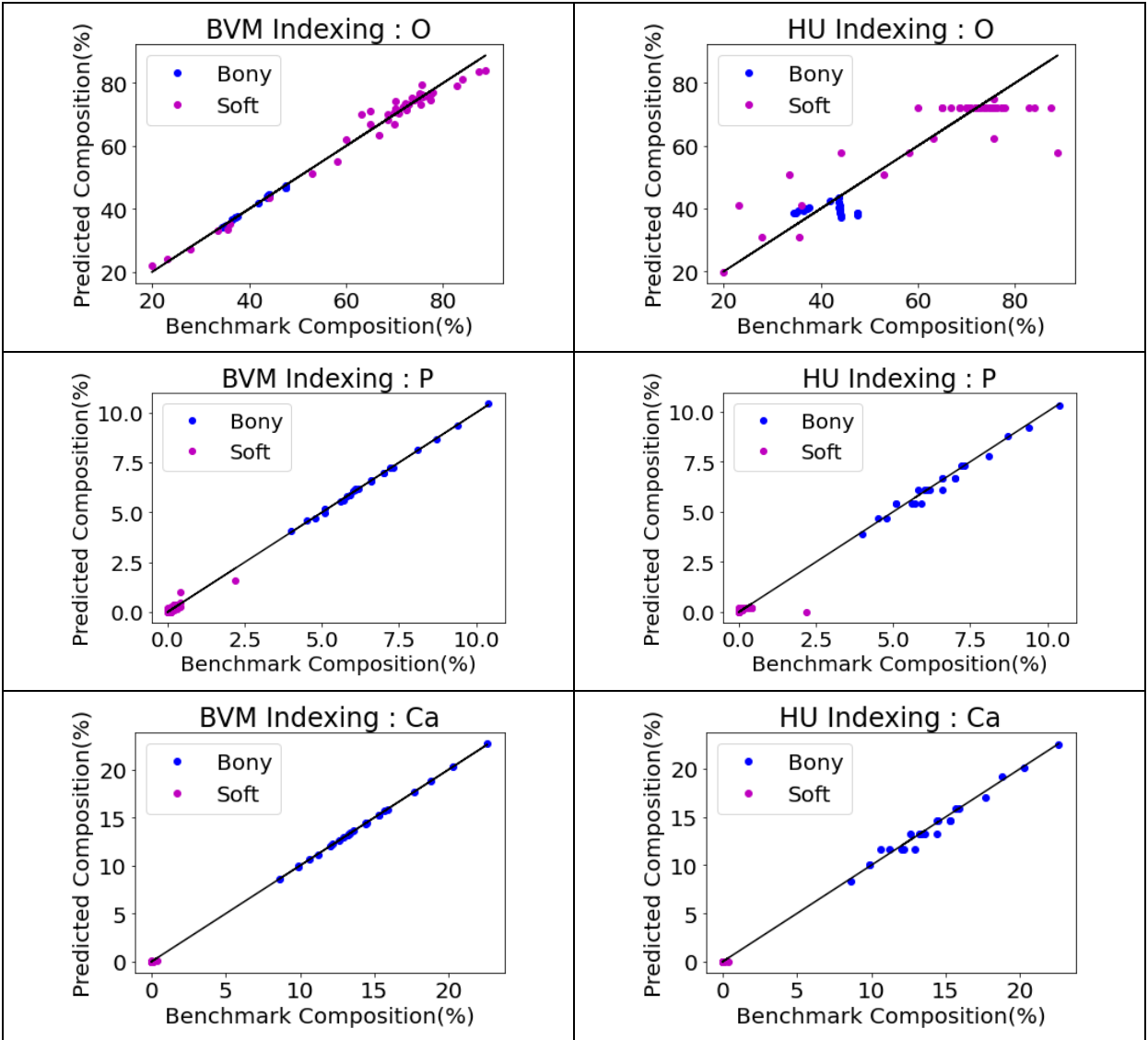

FIGURE. S1. Dispersion of the predicted BVM and HU material indexing atomic fraction for the 70 analyzed tissues. Straight lines denote perfect correspondence between benchmark and predicted composition percentages.

## S2. Compositional Properties of 70 Investigated Tissues

| | | | Soft Tissues | | | | | | |
|---|---|---|---|---|---|---|---|---|---|
| ID | Tissue | $\rho_m$ (g·cm$^{-3}$) | Composition(%) | | | | | | $[C_1, C_2]$ |
| | | | H | C | N | O | P | Ca | |
| 1 | Lunginflated | 0.26 | 10.4 | 10.6 | 3.1 | 75.7 | 0.2 | 0 | [0.22;0.03] |
| 2 | Adiposetissue3 | 0.93 | 11.6 | 68.3 | 0.2 | 19.9 | 0 | 0 | [0.89;0.02] |
| 3 | Adiposetissue2 | 0.95 | 11.4 | 60 | 0.7 | 27.9 | 0 | 0 | [0.89;0.03] |
| 4 | Adiposetissue1 | 0.97 | 11.2 | 51.9 | 1.3 | 35.6 | 0 | 0 | [0.90;0.04] |
| 5 | Yellowmarrow | 0.98 | 11.5 | 64.6 | 0.7 | 23.2 | 0 | 0 | [0.93;0.03] |
| 6 | Mammarygland1 | 0.99 | 10.9 | 50.8 | 2.3 | 35.9 | 0.1 | 0 | [0.91;0.05] |
| 7 | Yellowredmarrow | 1 | 11 | 53.2 | 2.1 | 33.6 | 0.1 | 0 | [0.92;0.04] |
| 8 | Mammarygland2 | 1.02 | 10.6 | 33.3 | 3 | 53 | 0.1 | 0 | [0.90;0.07] |
| 9 | Redmarrow | 1.03 | 10.6 | 41.7 | 3.4 | 44.2 | 0.1 | 0 | [0.93;0.06] |
| 10 | BrainCerebrospinalfluid | 1.01 | 11.2 | 0 | 0 | 88.8 | 0 | 0 | [0.85;0.12] |
| 11 | Adrenalgland | 1.03 | 10.7 | 28.5 | 2.6 | 58.1 | 0.1 | 0 | [0.91;0.08] |
| 12 | Smallintestinewall | 1.03 | 10.7 | 11.6 | 2.2 | 75.4 | 0.1 | 0 | [0.88;0.10] |

| ID | Tissue | $\rho_m$ (g·cm⁻³) | Composition(%) | | | | | | $[C_1, C_2]$ |
|----|--------|-------------------|----------------|---|---|---|---|---|--------------|
|    |        |                   | H | C | N | O | P | Ca |              |
| 13 | Urine | 1.02 | 11.1 | 0.5 | 1 | 87.3 | 0.1 | 0 | [0.86;0.12] |
| 14 | Gallbladderbile | 1.03 | 10.9 | 6.1 | 0.1 | 82.9 | 0 | 0 | [0.87;0.11] |
| 15 | Lymph | 1.03 | 10.9 | 4.1 | 1.1 | 83.9 | 0 | 0 | [0.87;0.11] |
| 16 | Pancreas | 1.04 | 10.7 | 17 | 2.2 | 69.9 | 0.2 | 0 | [0.89;0.10] |
| 17 | Brainwhitematter | 1.04 | 10.7 | 19.6 | 2.5 | 66.8 | 0.4 | 0 | [0.90;0.10] |
| 18 | Prostate | 1.04 | 10.6 | 9 | 2.5 | 77.8 | 0.1 | 0 | [0.88;0.11] |
| 19 | Testis | 1.04 | 10.7 | 10 | 2 | 77.2 | 0.1 | 0 | [0.89;0.11] |
| 20 | Braingraymatter | 1.04 | 10.8 | 9.6 | 1.8 | 77.5 | 0.3 | 0 | [0.88;0.11] |
| 21 | Muscleskeletal1 | 1.05 | 10.2 | 17.3 | 3.6 | 68.7 | 0.2 | 0 | [0.90;0.10] |
| 22 | Heart1 | 1.05 | 10.4 | 17.6 | 3.1 | 68.7 | 0.2 | 0 | [0.90;0.10] |
| 23 | Kidney1 | 1.05 | 10.3 | 16.1 | 3.4 | 69.9 | 0.2 | 0.1 | [0.89;0.11] |
| 24 | Stomach | 1.05 | 10.5 | 14 | 2.9 | 72.5 | 0.1 | 0 | [0.90;0.10] |
| 25 | Muscleskeletal2 | 1.05 | 10.3 | 14.4 | 3.4 | 71.7 | 0.2 | 0 | [0.90;0.10] |

| ID | Tissue | $\rho_m$ (g·cm$^{-3}$) | Composition(%) | | | | | | $[C_1, C_2]$ |
|---|---|---|---|---|---|---|---|---|---|
| | | | H | C | N | O | P | Ca | |
| 26 | Liver1 | 1.05 | 10.4 | 15.8 | 2.7 | 70.8 | 0.3 | 0 | [0.90;0.10] |
| 27 | Heart2 | 1.05 | 10.5 | 14 | 2.9 | 72.4 | 0.2 | 0 | [0.87;0.12] |
| 28 | Aorta | 1.05 | 10 | 14.8 | 4.2 | 70.2 | 0.4 | 0.4 | [0.90;0.10] |
| 29 | Kidney2 | 1.05 | 10.4 | 13.3 | 3 | 73 | 0.2 | 0.1 | [0.89;0.11] |
| 30 | Muscleskeletal3 | 1.05 | 10.3 | 11.3 | 3 | 75.2 | 0.2 | 0 | [0.89;0.11] |
| 31 | Heart3 | 1.05 | 10.5 | 10.4 | 2.7 | 76.2 | 0.2 | 0 | [0.89;0.11] |
| 32 | Mammarygland3 | 1.06 | 10.2 | 15.9 | 3.7 | 70.1 | 0.1 | 0 | [0.91;0.10] |
| 33 | Kidney3 | 1.05 | 10.5 | 10.7 | 2.7 | 75.8 | 0.2 | 0.1 | [0.89;0.11] |
| 34 | Ovary | 1.05 | 10.6 | 9.4 | 2.4 | 77.4 | 0.2 | 0 | [0.89;0.11] |
| 35 | Eyelens | 1.07 | 9.6 | 19.6 | 5.7 | 65 | 0.1 | 0 | [0.92;0.10] |
| 36 | Liver2 | 1.06 | 10.3 | 14 | 3 | 72.4 | 0.3 | 0 | [0.90;0.11] |
| 37 | Spleen | 1.06 | 10.4 | 11.4 | 3.2 | 74.7 | 0.3 | 0 | [0.90;0.11] |
| 38 | Trachea | 1.06 | 10.2 | 14 | 3.3 | 72.1 | 0.4 | 0 | [0.90;0.11] |

| ID | Tissue | $\rho_m$ (g·cm$^{-3}$) | Composition(%) | | | | | | $[C_1, C_2]$ |
|---|---|---|---|---|---|---|---|---|---|
| | | | H | C | N | O | P | Ca | |
| 39 | Heartbloodfilled | 1.06 | 10.4 | 12.2 | 3.2 | 74.1 | 0.1 | 0 | [0.90;0.11] |
| 40 | Bloodwhole | 1.06 | 10.3 | 11.1 | 3.3 | 75.2 | 0.1 | 0 | [0.90;0.11] |
| 41 | Liver3 | 1.07 | 10.2 | 12.7 | 3.3 | 73.5 | 0.3 | 0 | [0.91;0.11] |
| 42 | Skin1 | 1.09 | 10.1 | 25.2 | 4.6 | 60 | 0.1 | 0 | [0.95;0.09] |
| 43 | Skin2 | 1.09 | 10.1 | 20.6 | 4.2 | 65 | 0.1 | 0 | [0.94;0.10] |
| 44 | Skin3 | 1.09 | 10.2 | 15.9 | 3.7 | 70.1 | 0.1 | 0 | [0.93;0.10] |
| 45 | Connectivetissue | 1.12 | 9.5 | 21 | 6.3 | 63.2 | 0 | 0 | [0.96;0.10] |
| 46 | Cartilage | 1.1 | 9.8 | 10.1 | 2.2 | 75.7 | 2.2 | 0 | [0.88;0.15] |

| | | | Bony Tissues | | | | | | |
|---|---|---|---|---|---|---|---|---|---|
| ID | Tissue | $\rho_m$ (g·cm$^{-3}$) | Composition(%) | | | | | | $[C_1, C_2]$ |
| | | | H | C | N | O | P | Ca | |
| 1 | Sternum | 1.25 | 7.8 | 31.8 | 3.7 | 44.1 | 4 | 8.6 | [0.45;0.63] |
| 2 | Sacrummale | 1.29 | 7.4 | 30.4 | 3.7 | 44.1 | 4.5 | 9.9 | [0.36;0.74] |
| 3 | D6L3inclcartilagem | 1.3 | 7.4 | 26.7 | 3.6 | 47.6 | 4.8 | 9.9 | [0.35;0.75] |
| 4 | Vertcolwhole | 1.33 | 7.2 | 26 | 3.6 | 47.5 | 5.1 | 10.6 | [0.30;0.82] |
| 5 | VertcolD6L3exclcartilage | 1.33 | 7 | 28.9 | 3.8 | 44 | 5.1 | 11.2 | [0.27;0.85] |
| 6 | FemurHumerussphericalhead | 1.33 | 7.1 | 38.2 | 2.6 | 34.3 | 5.6 | 12.2 | [0.21;0.90] |
| 7 | Femurconicaltrochanter | 1.36 | 6.9 | 36.8 | 2.7 | 34.8 | 5.9 | 12.9 | [0.15;0.96] |
| 8 | C4inclcartilagemale | 1.38 | 6.6 | 24.5 | 3.7 | 47.5 | 5.7 | 12 | [0.19;0.95] |
| 9 | Sacrumfemale | 1.39 | 6.6 | 27.3 | 3.8 | 43.9 | 5.8 | 12.6 | [0.15;0.98] |
| 10 | Humeruswholespecimen | 1.39 | 6.7 | 35.4 | 2.8 | 35.3 | 6.2 | 13.6 | [0.09;1.04] |
| 11 | Ribs2nd6th | 1.41 | 6.4 | 26.5 | 3.9 | 44 | 6 | 13.2 | [0.11;1.04] |
| 12 | Innominatemale | 1.41 | 6.3 | 26.4 | 3.9 | 44 | 6.1 | 13.3 | [0.09;1.05] |

| ID | Tissue | $\rho_m$ (g·cm⁻³) | Composition(%) | | | | | | $[C_1, C_2]$ |
|---|---|---|---|---|---|---|---|---|---|
| | | | H | C | N | O | P | Ca | |
| 13 | VertcolC4exclcartilage | 1.42 | 6.3 | 26.3 | 3.9 | 44 | 6.1 | 13.4 | [0.09;1.07] |
| 14 | Femurtotalbone | 1.42 | 6.3 | 33.4 | 2.9 | 36.4 | 6.6 | 14.4 | [0.02;1.13] |
| 15 | Femurwholespecism | 1.43 | 6.3 | 33.2 | 2.9 | 36.5 | 6.6 | 14.5 | [0.01;1.14] |
| 16 | Innominatefemale | 1.46 | 6 | 25.2 | 3.9 | 43.9 | 6.6 | 14.4 | [-0.0025;1.17] |
| 17 | Humerustotalbone | 1.46 | 6 | 31.5 | 3.1 | 37.1 | 7 | 15.3 | [-0.06;1.23] |
| 18 | Claviclescapula | 1.46 | 6 | 31.4 | 3.1 | 37.2 | 7 | 15.3 | [-0.06;1.23] |
| 19 | Humeruscylindricalshaft | 1.49 | 5.8 | 30.3 | 3.2 | 37.6 | 7.2 | 15.9 | [-0.12;1.30] |
| 20 | Ribs10th | 1.52 | 5.6 | 23.7 | 4 | 43.7 | 7.3 | 15.7 | [-0.13;1.33] |
| 21 | Cranium | 1.61 | 5 | 21.3 | 4 | 43.9 | 8.1 | 17.7 | [-0.34;1.57] |
| 22 | Mandible | 1.68 | 4.6 | 20 | 4.1 | 43.8 | 8.7 | 18.8 | [-0.47;1.74] |
| 23 | Femurcylindricalshaft | 1.75 | 4.2 | 20.5 | 3.8 | 41.8 | 9.4 | 20.3 | [-0.65;1.94] |
| 24 | Corticalbone | 1.92 | 3.4 | 15.6 | 4.2 | 43.8 | 10.4 | 22.6 | [-0.99;2.36] |

## S3. Lateral dose profiles of three representative tissues for two depth definitions

Figure S2 presents lateral dose profile curves for representative tissues with the highest R80 discrepancies. Depth-dose profiles for each material indexing method were aligned based on the Bragg peak and R80 position of the ground truth composition. The corresponding lateral dose profiles at the Bragg peak and R80 positions were then extracted for comparison. The dashed line indicates the mean dose in the core region (100% to 10%) (yellow), the halo region (subdivided into proximal (10% to 1%) and distal (1% to 0.1%) components) (green), and the proximal aura region (0.1% to 0.01%) (blue).

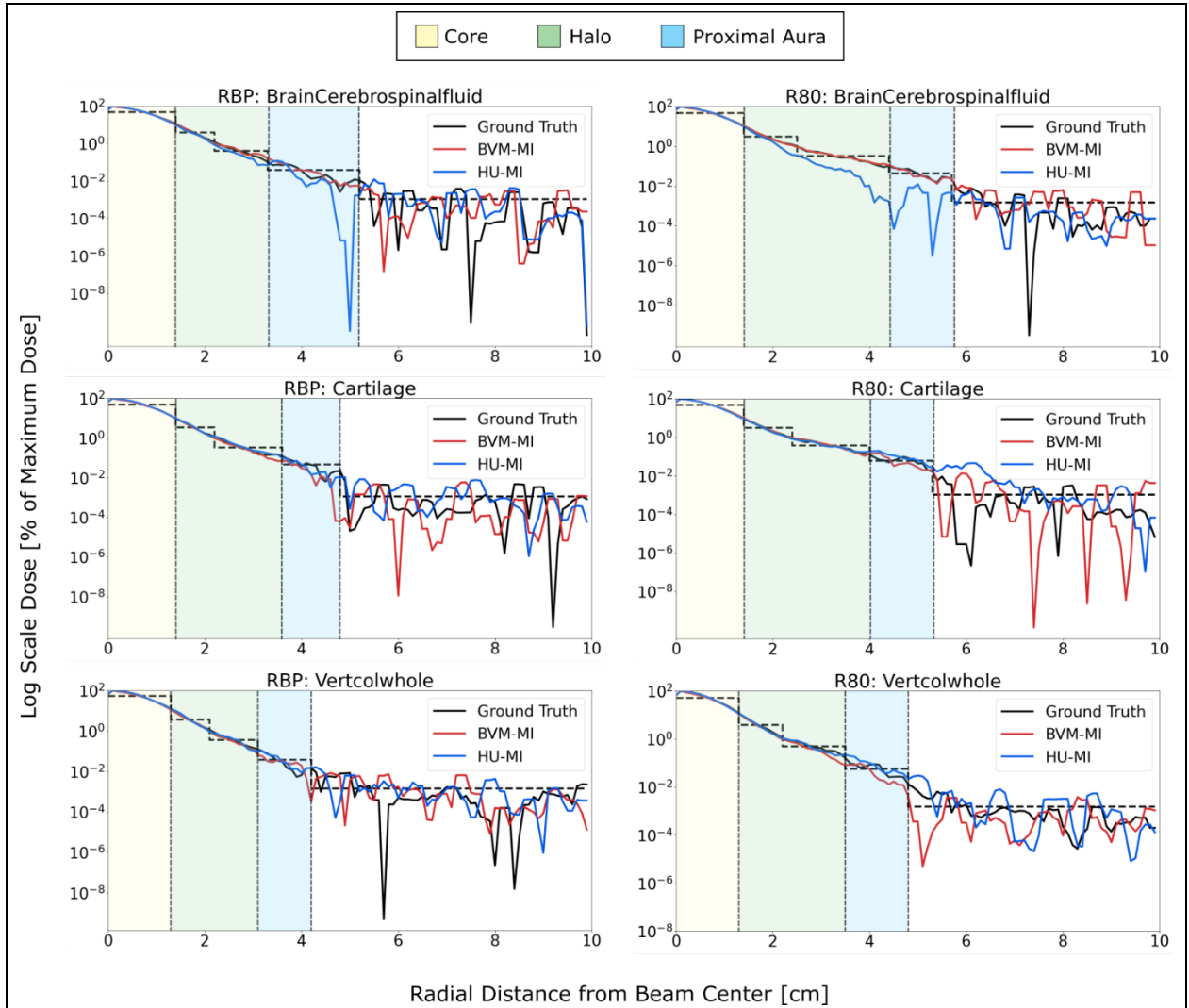

FIGURE. S2. Lateral dose profile curves relative to maximum dose at different radial distances for representative soft tissue (Brain Cerebrospinal Fluid and Cartilage) and bony tissue (Vertebral Column Whole) with the highest R80 discrepancies.

## S4. Lateral dose profiles of three representative tissues for two RBP depth definitions

TABLE S1. Mean Absolute Percent Error Between Doses in Core, Halo, and Proximal Aura Regions for Benchmark, BVM-MI, and HU-MI MC Simulations scored at Bragg peak range.

|  |  | Brain Cerebrospinal Fluid | | Cartilage | | Vertebral Column Whole | |
|---|---|---|---|---|---|---|---|
|  |  | **BVM-MI** | **HU-MI** | **BVM-MI** | **HU-MI** | **BVM-MI** | **HU-MI** |
| **Core** | $D_\% > 10\%$ | 1.8%* | 5.7% | 2.5% | 1.9% | 3.8% | 1.6% |
| **Halo** | $10\% < D_\% \leq 1\%$ | 8.5%* | 18.0% | 4.9% | 4.9% | 7.7% | 9.9% |
|  | $1\% \leq D_\% < 0.1\%$ | 26.0%* | 40.1% | 17.9%* | 15.6% | 10.2% | 17.0% |
| **Proximal Aura** | $0.1\% \leq D_\% < 0.01\%$ | 53.4%* | 56.1% | 54.7% | 40.7% | 60.2% | 19.9% |

*Statistically significant differences ($p < 0.05$) between BVM-MI and HU-MI results.

## S5. Dose to medium profiles along core, halo, and proximal aura regions

Figures S4–S6 present depth–dose profiles at radial distances corresponding to 10%, 1%, 0.1%, and 0.01% of the central-axis Bragg peak dose for three representative tissues exhibiting the largest R80 discrepancies among soft and bony tissues. The 10% distance corresponds to the core-halo region boundary and 0.01% marks the halo-aura boundary region.

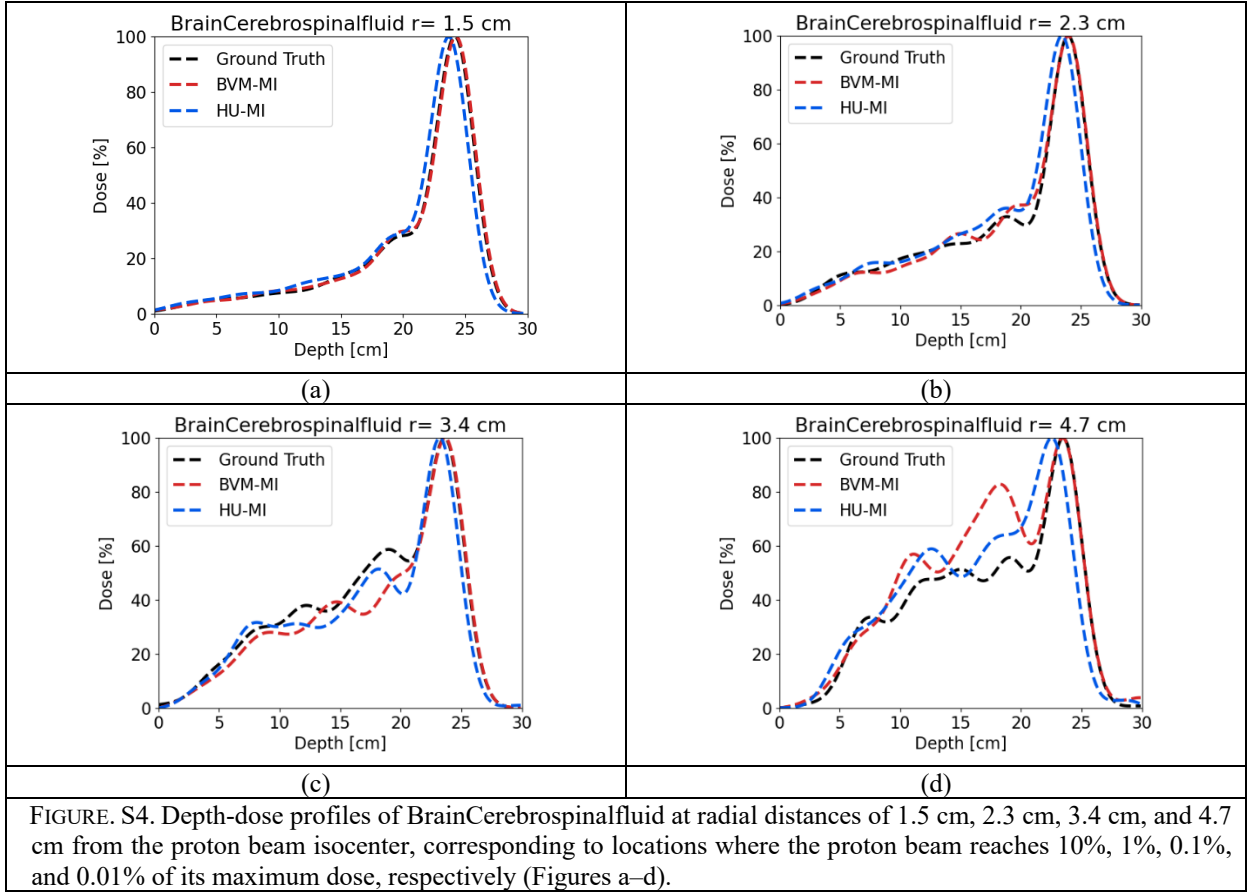

FIGURE. S4. Depth-dose profiles of BrainCerebrospinalfluid at radial distances of 1.5 cm, 2.3 cm, 3.4 cm, and 4.7 cm from the proton beam isocenter, corresponding to locations where the proton beam reaches 10%, 1%, 0.1%, and 0.01% of its maximum dose, respectively (Figures a–d).

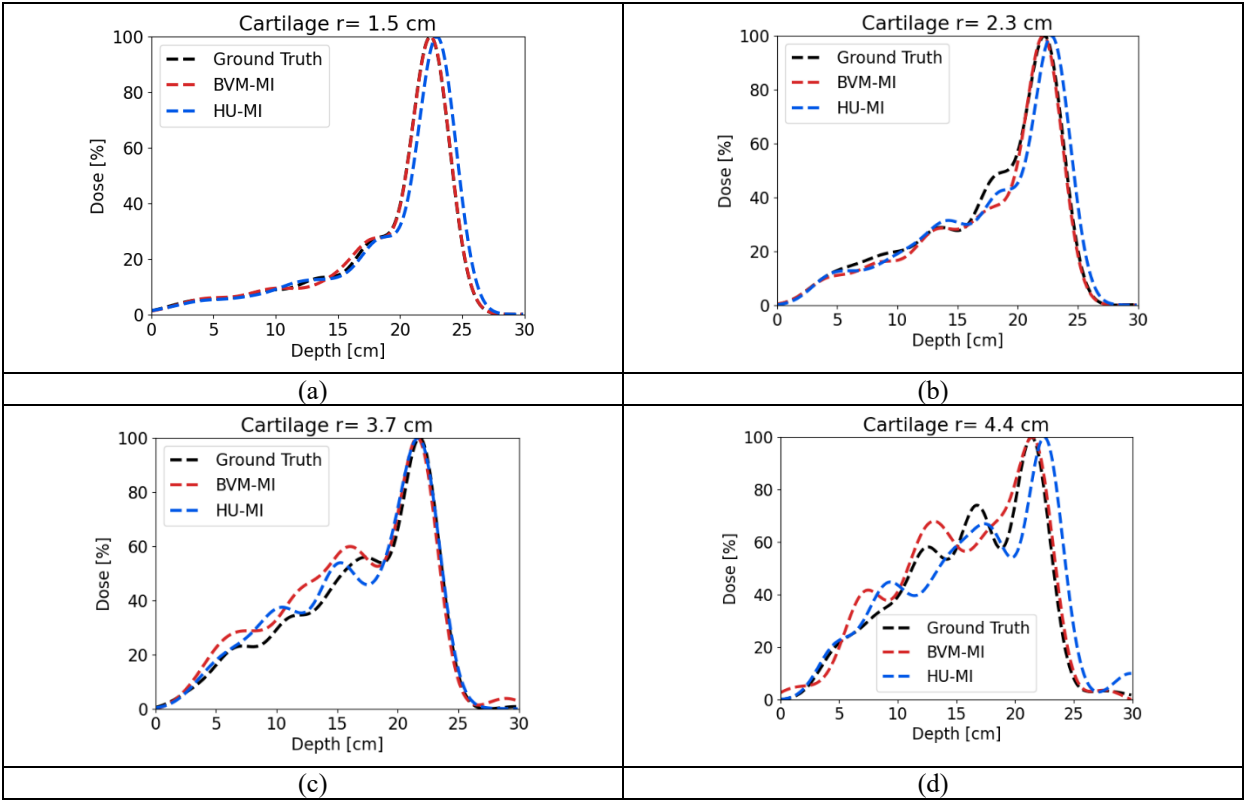

FIGURE. S5. Depth-dose profiles of Cartilage at radial distances corresponding to 10% (1.5 cm), 1% (2.3 cm), 0.1% (3.7 cm), and 0.01% (4.4 cm) of the maximum dose (Figures a–d).

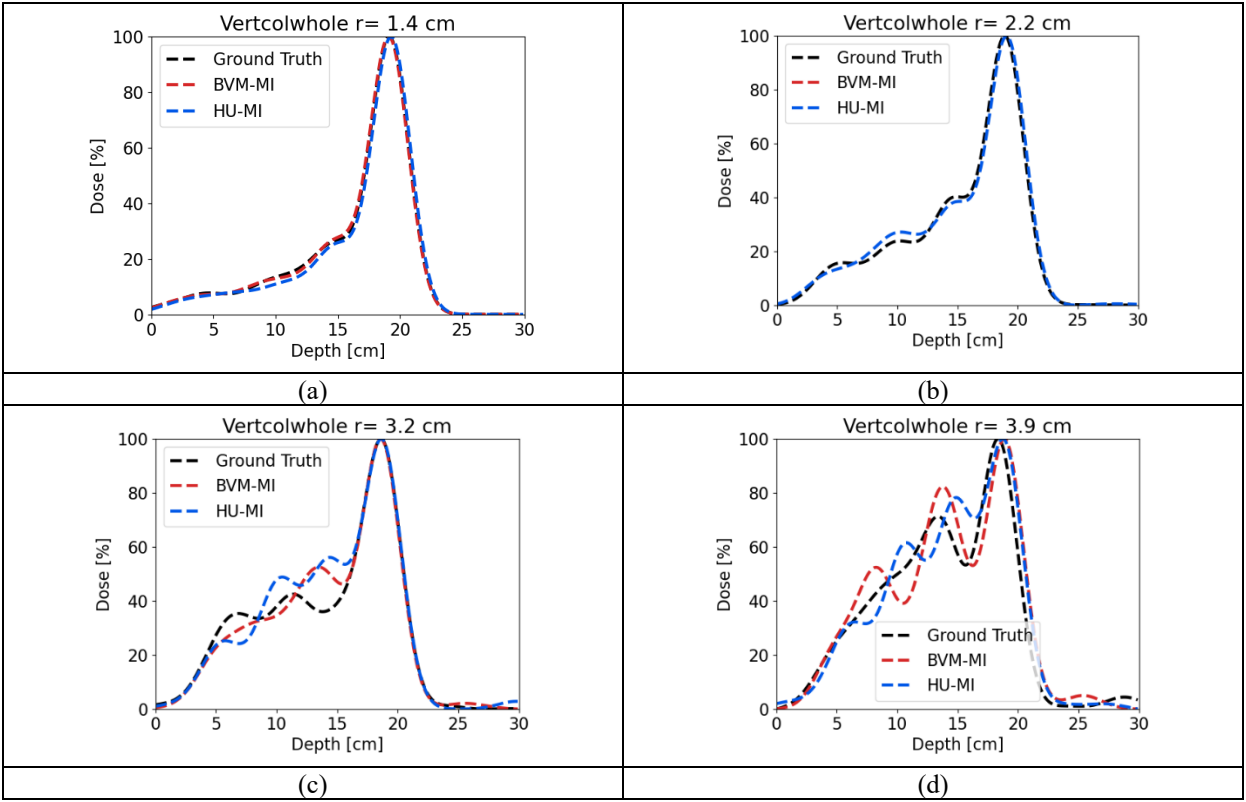

FIGURE. S6. Depth-dose profiles of Vertcolwhole at radial distances corresponding to 10% (1.4 cm), 1% (2.2 cm), 0.1% (3.2 cm), and 0.01% (3.9 cm) of the maximum dose (Figures a–d).

## S6. HU Material Indexing Method

The HU-based material-indexing (HU-MI) method characterizes tissue elemental composition and density using photon attenuation values measured in Hounsfield Units (HU) from SECT. To perform HU-MI material parametrization, each tissue is modeled as a combination of two primary material components. A parametrization function is then used to determine the tissue's density and elemental composition based on the HU and density of the bases material components and the CT number of the scanned tissue. Schneider et al. developed three parametrization functions to cover the HU ranges for soft and bony tissues. Using elemental composition data from 71 common reference tissues and these functions, a HU-to-material composition look-up table was created, consisting of 24 HU bins. In our implementation, we adopted Schneider's original 24-bin look-up table (Table S4), and mass densities were determined from the tissue's CT number by applying the original parametrization functions,

$$\begin{aligned} -22 \leq HU \leq 1524 & \quad \rho_m = (1.017 + 0.592 \times 10^{-3} H) g/cm^3, \\ -98 \leq HU \leq 14 & \quad \rho_m = (1.018 + 0.893 \times 10^{-3} H) g/cm^3, \\ 23 \leq HU \leq 100 & \quad \rho_m = (1.003 + 1.169 \times 10^{-3} H) g/cm^3, \end{aligned} \quad (1)$$

TABLE S2. Conversion from HU Value to elemental weight composition. For this study, we employed the original 24-bin conversion table developed by Schneider. (Schneider, 2000)

| HU | H | C | N | O | Na | Mg | P | S | Cl | Ar | K | Ca |
|---|---|---|---|---|---|---|---|---|---|---|---|---|
| [-1000, -950] | 0 | 0 | 75.5 | 23.2 | 0 | 0 | 0 | 0 | 0 | 1.3 | 0 | 0 |
| [-949, -120] | 10.3 | 10.5 | 3.1 | 74.9 | 0.2 | 0 | 0.2 | 0.3 | 0.3 | 0 | 0.2 | 0 |
| [-119, -83] | 11.6 | 68.1 | 0.2 | 19.8 | 0.1 | 0 | 0 | 0.1 | 0.1 | 0 | 0 | 0 |
| [-82, -53] | 11.3 | 56.7 | 0.9 | 30.8 | 0.1 | 0 | 0 | 0.1 | 0.1 | 0 | 0 | 0 |
| [-52, -23] | 11 | 45.8 | 1.5 | 41.1 | 0.1 | 0 | 0.1 | 0.2 | 0.2 | 0 | 0 | 0 |
| [-22, 7] | 10.8 | 35.6 | 2.2 | 50.9 | 0 | 0 | 0.1 | 0.2 | 0.2 | 0 | 0 | 0 |
| [8, 18] | 10.6 | 28.4 | 2.6 | 57.8 | 0 | 0 | 0.1 | 0.2 | 0.2 | 0 | 0.1 | 0 |
| [19, 80] | 10.3 | 13.4 | 3 | 72.3 | 0.2 | 0 | 0.2 | 0.2 | 0.2 | 0 | 0.2 | 0 |
| [81, 120] | 9.4 | 20.7 | 6.2 | 62.2 | 0.6 | 0 | 0 | 0.6 | 0.3 | 0 | 0 | 0 |
| [121, 200] | 9.5 | 45.5 | 2.5 | 35.5 | 0.1 | 0 | 2.1 | 0.1 | 0.1 | 0 | 0.1 | 4.5 |
| [201, 300] | 8.9 | 42.3 | 2.7 | 36.3 | 0.1 | 0 | 3 | 0.1 | 0.1 | 0 | 0.1 | 6.4 |
| [301, 400] | 8.2 | 39.1 | 2.9 | 37.2 | 0.1 | 0 | 3.9 | 0.1 | 0.1 | 0 | 0.1 | 8.3 |
| [401, 500] | 7.6 | 36.1 | 3 | 38 | 0.1 | 0.1 | 4.7 | 0.2 | 0.1 | 0 | 0 | 10.1 |
| [501, 600] | 7.1 | 33.5 | 3.2 | 38.7 | 0.1 | 0.1 | 5.4 | 0.2 | 0 | 0 | 0 | 11.7 |
| [601, 700] | 6.6 | 31 | 3.3 | 39.4 | 0.1 | 0.1 | 6.1 | 0.2 | 0 | 0 | 0 | 13.2 |
| [701, 800] | 6.1 | 28.7 | 3.5 | 40 | 0.1 | 0.1 | 6.7 | 0.2 | 0 | 0 | 0 | 14.6 |
| [801, 900] | 5.6 | 26.5 | 3.6 | 40.5 | 0.1 | 0.2 | 7.3 | 0.3 | 0 | 0 | 0 | 15.9 |
| [901, 1000] | 5.2 | 24.6 | 3.7 | 41.1 | 0.1 | 0.2 | 7.8 | 0.3 | 0 | 0 | 0 | 17 |
| [1001, 1100] | 4.9 | 22.7 | 3.8 | 41.6 | 0.1 | 0.2 | 8.3 | 0.3 | 0 | 0 | 0 | 18.1 |
| [1101, 1200] | 4.5 | 21 | 3.9 | 42 | 0.1 | 0.2 | 8.8 | 0.3 | 0 | 0 | 0 | 19.2 |
| [1201, 1300] | 4.2 | 19.4 | 4 | 42.5 | 0.1 | 0.2 | 9.2 | 0.3 | 0 | 0 | 0 | 20.1 |
| [1301, 1400] | 3.9 | 17.9 | 4.1 | 42.9 | 0.1 | 0.2 | 9.6 | 0.3 | 0 | 0 | 0 | 21 |
| [1401, 1500] | 3.6 | 16.5 | 4.2 | 43.2 | 0.1 | 0.2 | 10 | 0.3 | 0 | 0 | 0 | 21.9 |
| [1501, 1600] | 3.4 | 15.5 | 4.2 | 43.5 | 0.1 | 0.2 | 10.3 | 0.3 | 0 | 0 | 0 | 22.5 |

**REFERENCES**

Schneider, W.T.B. and W.S. (2000) 'Correlation between CT numbers and tissue parameters needed for Monte Carlo simulations of clinical dose distributions', *Phys. Med. Biol*, 45(2), pp. 459–478.